\documentclass[aps,prl,10pt,twocolumn,superscriptaddress,floatfix,noeprint]{revtex4-2}

\usepackage{graphicx}
\usepackage{amsmath,amssymb}
\usepackage{bm,dsfont}
\usepackage[dvipsnames]{xcolor}
\usepackage[T1]{fontenc}
\usepackage[utf8]{inputenc}
\usepackage{CJKutf8}
\usepackage[colorlinks=true,citecolor=Green,linkcolor=BrickRed,urlcolor=NavyBlue,bookmarks=false]{hyperref}

\DeclareMathOperator{\sgn}{sgn}
\DeclareMathOperator{\sech}{sech}

\def\unit#1{\mathord{\thinspace\rm #1}}
\def\func#1{\mathop{\rm #1}\nolimits}

\graphicspath{{figures/}}

\newcommand{\mytitle}{Gate-Tunable Resonances and 1D Channel in a Graphene Nanoslide}

\begin{document}
\begin{CJK}{UTF8}{bsmi}

\title{\mytitle}

\author{Christophe De Beule}
\email{christophe.debeule@uantwerpen.be}
\affiliation{Department of Physics and Astronomy, University of Pennsylvania, Philadelphia, Pennsylvania 19104, USA}
\affiliation{Department of Physics, University of Antwerp, Groenenborgerlaan 171, 2020 Antwerp, Belgium}

\author{Ming-Hao Liu (劉明豪)}
\email{minghao.liu@phys.ncku.edu.tw}
\affiliation{Department of Physics and Center for Quantum Frontiers of Research and Technology (QFort), National Cheng Kung University, Tainan 70101, Taiwan}

\author{Bart Partoens}
\affiliation{Department of Physics, University of Antwerp, Groenenborgerlaan 171, 2020 Antwerp, Belgium}

\author{Lucian Covaci}
\affiliation{Department of Physics and NANOlight Center of Excellence, University of Antwerp, Groenenborgerlaan 171, 2020 Antwerp, Belgium}

\date{\today}

\begin{abstract}
We present a theory of the graphene nanoslide, a fundamental device for graphene straintronics that realizes a single pseudogauge barrier. We solve the scattering problem in closed form and demonstrate that the nanoslide gives rise to a hybrid pseudogauge and electrostatic cavity in the bipolar regime, and hosts one-dimensional transverse channels. The latter can be tuned using a bottom gate between valley-chiral or counterpropagating modes, as well as one-dimensional flatbands. Hence, the local density of states near the barrier depends strongly on the gate voltage with a tunable sublattice and electron-hole asymmetry. In the presence of electron-electron interactions, the nanoslide allows for \textit{in-situ} tuning between a chiral and ordinary Tomonaga-Luttinger liquid.
\end{abstract}

\maketitle

\end{CJK}

Graphene straintronics offers a promising route for designing next-generation devices through the manipulation of strain-induced pseudomagnetic fields. In graphene, elastic shear deformations of the honeycomb lattice couple to the low-energy Dirac electrons as pseudogauge fields \cite{kane_size_1997,suzuura_phonons_2002,katsnelson_graphene_2007,vozmediano_gauge_2010}. For certain strain configurations, this gives rise to pseudomagnetic fields \cite{guinea_generating_2010,guinea_energy_2010,low_gaps_2011} with opposite signs in the two valleys to preserve time-reversal symmetry, and with magnitudes that can exceed several hundreds of Tesla \cite{levy_strain-induced_2010,nigge_room_2019}. Similar pseudogauge fields also arise in strained semiconductors \cite{iordanskii_dislocations_1985,roldan_strain_2015,cazalilla_quantum_2014,rostami_theory_2015} and topological semimetals \cite{cortijo_elastic_2015,pikulin_chiral_2016,ilan_pseudo-electromagnetic_2020}. For example, substrate-engineered periodic corrugations \cite{venderbos_interacting_2016,banerjee_strain_2020} give rise to a nonlinear anomalous Hall effect in bilayer graphene \cite{ho_hall_2021} and provide an alternative platform for engineering topological flatbands \cite{mao_evidence_2020,milovanovic_band_2020,  phong_boundary_2022,de_beule_network_2023} and strongly correlated electronic phases \cite{manesco_correlations_2020,manesco_correlation-induced_2021,gao_untwisting_2023} in monolayer graphene.
\begin{figure}
    \centering
    \includegraphics[width=\linewidth]{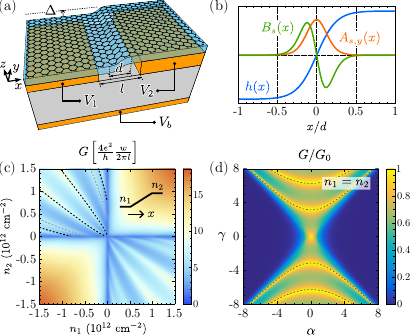}
    \caption{\textbf{Graphene nanoslide.} (a) Illustration of the graphene nanoslide device for the transport axis $x$ along the armchair lattice direction. (b) Height profile, pseudogauge barrier, and pseudomagnetic dipole of the nanoslide. (c) Two-terminal conductance as a function of carrier density in the leads for $V_b = 0$. Calculated with the transfer-matrix method for a linear density profile (see inset) with $\alpha = 0.8$, $l = 100$~nm, and $n_i = -V_i |V_i| / ( \pi v_F^2 )$ ($i=1,2$) the density in the leads. The dashed lines follow from Eq.\ \eqref{eq:interference} for $\delta = 0.16$ and match the resonances in the bipolar regime ($n_1/n_2 < 0$). (d) Two-terminal conductance for $n_1 = n_2$ in units of the ballistic conductance as a function of the pseudogauge [$\alpha \sim \Delta^2 / (da_0)$] and electrostatic [$\gamma \sim V_b l / v_F$] barrier.}
    \label{fig:setup}
\end{figure}

Here, we present a theory of the \textit{nanoslide}, which realizes a \textit{single} one-dimensional (1D) pseudogauge barrier by suspending monolayer graphene between two vertically-misaligned gates. In particular, we argue that recently observed oscillations in the two-terminal resistance \cite{zhang_gate-tunable_2022} originate from a hybrid interferometer, consisting of a central strain barrier and a moving PN junction rather than a Veselago lens. While previous works have considered interfaces between regions of different constant strain \cite{wu_valley-dependent_2011,wang_valley-polarized_2023} or nanowrinkles \cite{wu_quantum_2018,jun_nanowrinkle_2025}, the nanoslide is more elementary, as each region or wrinkle can be decomposed into two slides. Our work also provides insights into the scattering mechanism at substrate steps \cite{nakatsuji_uniaxial_2012,low_deformation_2012,
baringhaus_exceptional_2014,banerjee_strain_2020}. This Letter is organized as follows. We first introduce the graphene nanoslide and the resulting pseudogauge field. We then consider a two-terminal setup and compute the scattering matrix, bound states, and conductance in closed form for constant electron density, and verify these results with tight-binding simulations. In the bipolar regime, the nanoslide hosts a hybrid pseudogauge and electrostatic cavity, and the conductance agrees remarkably well with experiment \cite{zhang_gate-tunable_2022}. In addition, we consider a tunnel barrier from a bottom gate resulting in conductance resonances when bound states, localized at the barrier, merge with the Dirac continuum. These bound states form a 1D transverse channel that can be tuned between valley-chiral and counterpropagating electron or hole modes. Finally, we consider the sublattice-resolved local density of states which displays a gate-tunable sublattice and electron-hole asymmetry. 

\textcolor{NavyBlue}{\textit{Graphene nanoslide}} --- To realize a single pseudogauge barrier, we consider monolayer graphene subject to a 1D monotonic out-of-plane displacement field $h(x)$ with $h(x \rightarrow \pm \infty) = \pm \Delta/2$. For example,
\begin{equation} \label{eq:h}
    h(x) = (\Delta/2) \tanh \left( 4 x / d \right),
\end{equation}
where $d$ is the width over which $h(x)$ varies appreciably, relative to the net displacement $\Delta$. This models the experimental profile of Ref.\ \cite{zhang_gate-tunable_2022} where a graphene ribbon was suspended between two vertically misaligned gate electrodes, as illustrated in Fig.\ \ref{fig:setup}(a).

Such a height profile gives rise to uniaxial strain whose shear component breaks the microscopic $\mathcal C_{3z}$ symmetry about a carbon atom. Consequently, shear deformations that are smooth on the atomic scale, couple to the Dirac electrons in graphene as intravalley pseudogauge fields \cite{katsnelson_graphene_2007,vozmediano_gauge_2010}. In particular, the low-energy continuum Hamiltonian at valley $K$ and $K'$ is given by \cite{divincenzo_self-consistent_1984, castro_neto_electronic_2009}
\begin{equation}
    H = v_F \left[ -i \nabla_{\bm r} + e \bm A_\mathrm{tot}(\bm r) \right] \cdot (\tau_z \sigma_x, \sigma_y) + V(\bm r) \sigma_0,
\end{equation}
with $v_F \approx 10^6$~m~s$^{-1}$ the Fermi velocity of graphene, $\sigma_{x,y}$ ($\tau_z$) sublattice (valley) Pauli matrices, and we set $\hbar = 1$. Here $V(\bm r)$ is the electrostatic potential due to the gate electrodes, and $\bm A_\text{tot} = \bm A + \tau_z \bm A_s$ with $\bm A$ the vector potential from an external magnetic field and
\begin{equation}
    \bm A_s(\bm r) = \frac{\beta}{2ea_0} R(3\theta) \begin{pmatrix} u_{yy} - u_{xx} \\ u_{xy} + u_{yx} \end{pmatrix},
\end{equation}
the pseudogauge field due to strain \cite{vozmediano_gauge_2010}. Here $a_0 \approx 1.42$~{\r A} is the graphene carbon-carbon distance, $e$ the elementary charge, and $\theta$ the angle between the $x$ axis and the zigzag direction of the graphene lattice. We take $\beta$ as a phenomenological parameter to account for in-plane relaxation, similar to a reduction factor \cite{suzuura_phonons_2002}. See Supplementary Material (SM) \footnote{See Supplemental Material at [insert url] for details on the calculation of the scattering matrix and the effect of lattice relaxation.} and Refs.\ \cite{guinea_gauge_2008,wehling_midgap_2008} therein. For a 1D height profile,
\begin{equation}
    \bm A_s(x) = -\frac{\beta}{4ea_0} \left( \frac{dh}{dx} \right)^2 \begin{pmatrix} \cos 3\theta \\ \sin 3\theta \end{pmatrix},
\end{equation}
giving rise to a pseudomagnetic field (PMF)
\begin{equation} \label{eq:pmf}
    B_s = \partial_x A_{s,y} - \partial_y A_{s,x} = -\frac{\beta}{2ea_0} \frac{dh}{dx} \frac{d^2h}{dx^2} \sin(3\theta),
\end{equation}
in the $z$ direction, as shown in Fig.\ \ref{fig:setup}(b) for Eq.\ \eqref{eq:h}. Note the threefold-symmetric dependence on the lattice orientation relative to the uniaxial strain. In particular, for strain along the zigzag direction ($\theta = 0~\mathrm{mod}~\pi/3$) $\bm A_s$ is pure gauge and the PMF vanishes. In the remainder of this Letter, we use a transverse gauge $A_{s,x} = 0$.

We first consider the pseudogauge barrier in the absence of external potentials. When the strain is confined to a region much smaller than the Fermi wavelength, but still varies slowly on the lattice scale ($a \ll d \ll \lambda_F$) we can let $\bm A_s(x) \rightarrow \left( \alpha / e \right) \delta(x) \bm e_y$
where $\alpha$ is a dimensionless constant. One can assign the integrated weight $\alpha = -\beta \Delta^2 \sin(3\theta) / (3a_0 d)$ or take it as a new parameter whose physical meaning becomes clear later.
This is a good approximation for the density regime in experiments \cite{zhang_gate-tunable_2022}. Indeed, taking into account spin and valley, we have $k_F = \sqrt{\pi|n|}$ with $n$ the electron density, and we find $\lambda_F = 2\pi / k_F \approx 35 \text{ nm} \sqrt{10^{12} \text{ cm}^{-2} / |n|}$.

We now solve the scattering problem. Using translational symmetry, we obtain an effective Hamiltonian for each transverse mode and valley:
\begin{equation}
    \mathcal H(x,k_y) = -i v_F \tau \sigma_x \frac{d}{dx} + e v_F \Lambda^\mu(x,k_y) \sigma_\mu,
\end{equation}
with $\tau = \pm 1$, $\sigma_\mu = \left( \sigma_0, \, \sigma_x, \, \sigma_y, \, \sigma_z \right)$, and $\Lambda^\mu(x, k_y)$ a generalized potential. The wave equation is solved by
\begin{equation}
    \psi(x) = \mathcal P \exp \left[ - ie\tau \int_{x_0}^x dx' \Lambda^\mu(x',k_y) \sigma_x\sigma_\mu \right] \psi(x_0),
\end{equation}
where $\mathcal P$ is path-ordering \cite{timm_transport_2012} and we absorbed the energy in the potential.
Using this solution for a pseudogauge delta barrier, we find
\begin{equation} \label{eq:boundary}
    \psi(0^+) = e^{\alpha \sigma_z} \psi(0^-),
\end{equation}
such that the Dirac spinor is discontinuous at $x=0$ because of the delta barrier. An alternative derivation of Eq.\ \eqref{eq:boundary} is given in the SM \cite{Note1}. On the left-hand ($x<0$) and right-hand ($x>0$) sides of the barrier, there are no potentials and the scattering solution can be written as
\begin{equation} \label{eq:sol}
    \psi_j(x) = \frac{a_j}{\sqrt{2}} \begin{pmatrix} 1 \\ \tau e^{is\tau \phi} \end{pmatrix} e^{isqx} + \frac{b_j}{\sqrt{2}} \begin{pmatrix} 1 \\ -\tau e^{-is\tau \phi} \end{pmatrix} e^{-isqx},
\end{equation}
where $j = L, R$, $s = \sgn(E)$, and $q = k_F \cos \phi > 0$ with $\phi \in [-\pi/2,\pi/2]$ the incident angle and $k_F = |E/v_F|$. The sign ensures that electrons ($s=1$) and holes ($s=-1$) propagate in the correct direction. Our goal is to find the scattering matrix relating incoming and outgoing modes: $\left(b_L, \, a_R \right) = S \left( a_L, \, b_R \right)$. Imposing boundary conditions for the incident wave and from Eq.\ \eqref{eq:boundary} gives,
\begin{widetext}
\begin{equation}
    S(k_y,E) = \begin{pmatrix} r & t \\ t' & r' \end{pmatrix} = \frac{1}{q\cosh \alpha + i s \tau k_y \sinh \alpha} \begin{bmatrix} -\left( q + i s \tau k_y \right) \sinh \alpha & q \\ q & \left( q - i s \tau k_y \right) \sinh \alpha \end{bmatrix}.
\end{equation}
\end{widetext}

The form of the $S$ matrix with $t=t'$ and $|r|=|r'|$ follows from our gauge choice. Indeed, in the transverse gauge, there is an effective time-reversal symmetry
\begin{equation}
    \left( \sigma_z \mathcal K \right) \mathcal H(x,k_y) \left( \sigma_z \mathcal K \right)^{-1} = \mathcal H(x,k_y),
\end{equation}
for fixed valley and $k_y$ with $\mathcal K$ complex conjugation, which is manifest in Eq.\ \eqref{eq:sol}. There is also a mirror symmetry: $\sigma_y \mathcal H(-x,k_y) \sigma_y = \mathcal H(x,k_y)$. However, it is not manifest in our solution such that $r$ and $r'$ differ by a phase, see SM \cite{Note1}.
The transmission function becomes
\begin{equation} \label{eq:T}
    T(\phi) = |t|^2 = \frac{\cos^2 \phi}{\cosh^2 \alpha - \sin^2 \phi},
\end{equation}
and thus Klein tunneling is absent: $T(0) = \sech^2 \alpha \leq 1$ because pseudospin conservation is broken by the boundary condition. This gives a physical interpretation of $\alpha$ in terms of the barrier transparency.
From the poles of the $S$ matrix under analytic continuation $q \rightarrow is\kappa$, we find valley-chiral bound states with dispersion
\begin{equation} \label{eq:bound}
    E_b = \pm ( v_F \sech \alpha ) k_y, \quad \text{for} \quad \tau k_y \alpha < 0,
\end{equation}
and wave function
\begin{equation} \label{eq:boundwave}
    \psi_b(x) = a e^{-\kappa |x|} \left[ \Theta(-x) \begin{pmatrix} 1 \\ \pm i e^\alpha \end{pmatrix} + \Theta(x) \begin{pmatrix} e^\alpha \\ \pm i \end{pmatrix} \right],
\end{equation}
with $a = \sqrt{(1-\tanh \alpha) \kappa / 2}$, $\kappa = k_F |\sinh \alpha|$, and $\Theta(x)$ the Heaviside step function. These are interpreted as pseudomagnetic snake states trapped by the PMF dipole \cite{ho_hall_2021,liu_snake_2015}. Hence, a graphene nanoslide hosts a 1D valley-chiral channel. Integrating Eq.\ \eqref{eq:T} over all incident angles, we find the two-terminal conductance
\begin{equation} \label{eq:g2}
    G/G_0 = 1 + \sinh \alpha \tanh \alpha \ln \left| \tanh \frac{\alpha}{2} \right|,
\end{equation}
with $G_0 =(4e^2/h) w k_F / \pi$ the ballistic value for a ribbon of width $w$ with periodic boundary conditions. This formula agrees well with tight-binding simulations (see SM \cite{Note1} for details and also Ref.\ \cite{chakraborti_electron_2024} therein) as long as $\lambda_F \gg d$, as shown in Fig.\ \ref{fig:tb}(a) and (b).
\begin{figure}
    \centering
    \includegraphics[width=\linewidth]{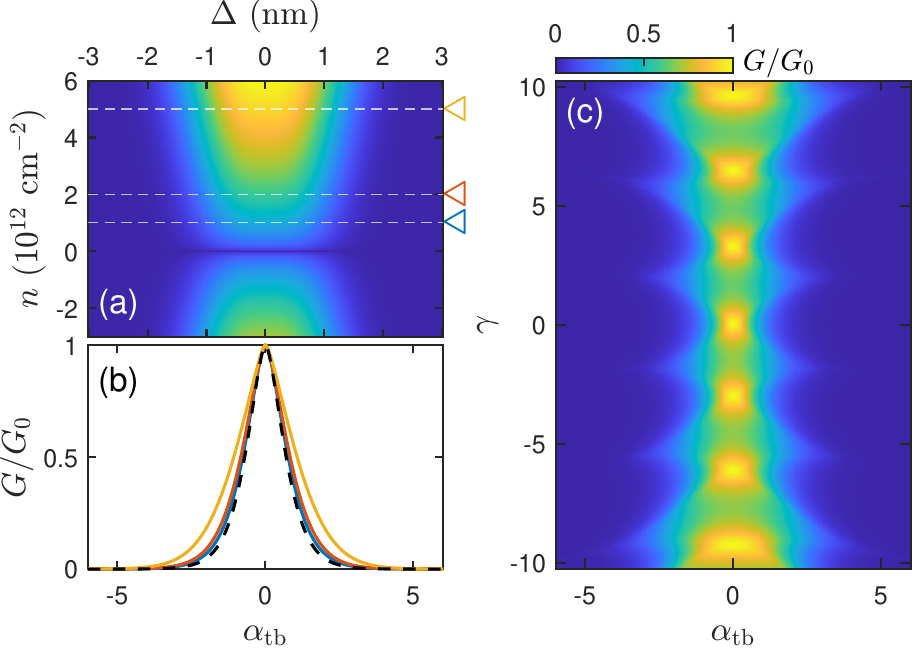}
    \caption{\textbf{Tight-binding simulations.} Two-terminal conductance calculated with tight binding for $x \parallel$ armchair, using the height profile \eqref{eq:h} for $\Delta = 1$~nm and $d = 10$~nm. We use the hopping function $t(r) = -t_0\exp \left[ \beta(1 - r/a_0) \right]$ with $t_0 = 3$~eV and $\beta = 3.37$ \cite{pereira_tight-binding_2009}. (a) Only the strain barrier as a function of density $n$ and $\alpha_\text{tb} \equiv \beta \Delta |\Delta| / (3a_0d)$. (b) Line cuts marked in (a). Deviations from Eq.\ \eqref{eq:g2} (dashed line) for $n > 10^{12}$~cm$^{-2}$ appear when $\lambda_F$ becomes comparable to the barrier width $d$. (c) With a potential $V(x) = V_b \sech^4(4x/l)$ due to a bottom gate for $l = 50$~nm where $\gamma = V_bl / (3v_F)$ and $n = 10^9$~cm$^{-2}$.} 
    \label{fig:tb}
\end{figure}

We have solved the low-energy scattering problem for the graphene nanoslide when the density in the leads $n_1 = n_2 = s k_F^2/\pi$. For a general profile $n(x)$ we have to resort to numerical methods \cite{Note1}. In particular, in the bipolar regime ($n_1/n_2 < 0$) one obtains a hybrid cavity with interference oscillations given by a quantization rule
\begin{equation} \label{eq:interference}
    \int_0^{x_0(n_1,n_2)} dx \, \sqrt{\pi \left| n(x) \right|} = \pi \left( m \pm \delta \right), \quad m \in \mathds Z,
\end{equation}
where $x_0$ is the charge neutrality point and $\delta$ gives a phase shift from reflection at the barriers ($\pm$ for time-reversed paths). For a linear density profile, Eq.\ \eqref{eq:interference} gives $|n_1 + n_2|^{3/2} = 3\sqrt{2 \pi} \left| n_1 - n_2 \right| (m \pm \delta) / l$ with $l$ the distance between the gates that set the local carrier density. The two-terminal conductance $G(n_1,n_2)$ is shown in Fig.\ \ref{fig:setup}(c) and agrees well with experiment \cite{zhang_gate-tunable_2022}. Note that for $x$ along the zigzag direction, the PMF vanishes ($\alpha = 0$) and there is no interference effect.
\begin{figure}
    \centering
    \includegraphics[width=\linewidth]{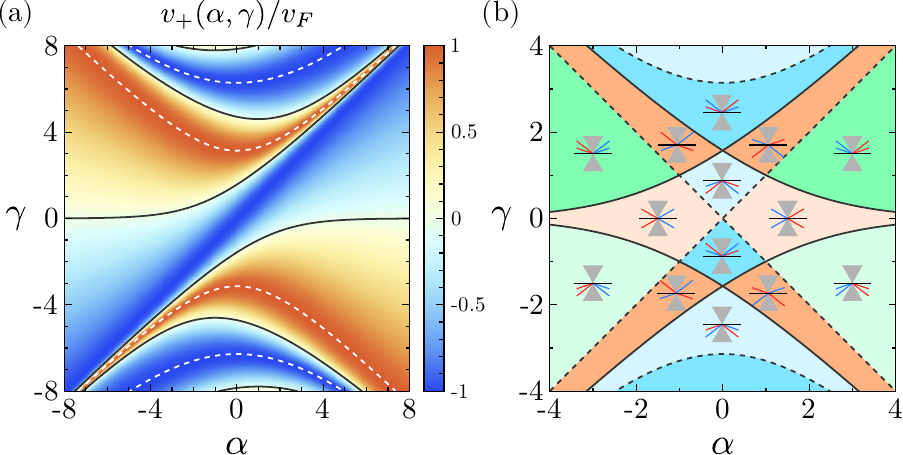}
    \caption{\textbf{Gate-tunable 1D channel.} (a) Velocity $v_+(\alpha,\gamma)$ of bound states with dispersion $E_b = v_\pm k_y$ with $v_-(\alpha,\gamma) = -v_+(\alpha,-\gamma)$. Solid and dashed curves give $v_+ = 0$ and $|v_+| = v_F$, respectively. (b) Phase diagram: cyan (light cyan) regions host counterpropagating electron (hole) modes in both valleys; orange (light orange) regions support one valley-chiral mode with equal (opposite) chirality for electrons and holes; and green (light green) regions host two valley-chiral modes for electrons (holes). In each region, a typical bound-state dispersion is sketched for valley $K$ (blue) and $K'$ (red), where the gray cone gives the bulk continuum.}
    \label{fig:channel}
\end{figure}

\textcolor{NavyBlue}{\textit{Gate-tunable resonances and 1D channel}} --- To further tune the properties of the nanoslide, we consider an additional tunnel barrier due to a bottom gate; see Fig.\ \ref{fig:setup}(a). The resulting potential is screened by the metal leads apart from the suspended region, which naturally aligns with the pseudogauge barrier. Assuming that the potential width is small relative to $\lambda_F$ we can use $V(x) = v_F \gamma \delta(x)$. As shown in the SM \cite{Note1}, the scattering problem can still be solved analytically. The resulting transmission at normal incidence is now given by $T(0) = \zeta^2 / \left( \alpha^2 \cosh^2 \zeta - \gamma^2 \right)$ with $\zeta = (\alpha^2 - \gamma^2)^{1/2}$ and the two-terminal conductance is shown in Fig.\ \ref{fig:setup}(d). Interestingly, it displays gate-tunable resonances for $\gamma^2 = \alpha^2 + (m\pi)^2$ ($m=1,2,\ldots$) that are qualitatively reproduced by tight-binding simulations, see Fig.\ \ref{fig:tb}(c). These resonances are due to bound states merging with the continuum ($\kappa \rightarrow 0^+$) that disperse linearly with velocity
\begin{equation}
    \frac{v_\pm}{v_F} = \frac{\pm \tau \alpha \sech \zeta - \gamma}{\tau \alpha \mp \gamma \sech \zeta} = \left< \sigma_y \right>_\pm,
\end{equation}
as  shown in Fig.\ \ref{fig:channel}(a), and inverse decay length $\kappa_\pm = k_F \sqrt{(v_F/v_\pm)^2 - 1}$. Here, the equality on the right-hand side follows from the Hellmann-Feynman theorem with respect to $k_y$. Note that the pseudospin points along $\sigma_y$ if the bound state merges with the continuum, as expected for a bulk state with momentum $(0, k_y)$. We also see that we recover Eq.\ \eqref{eq:bound} for $\gamma = 0$, and $v_\pm = \pm v_F \cos \gamma$ for $\alpha = 0$ with the condition $E \tan \gamma < 0$ \cite{yokoyama_gate-controlled_2010}.
Hence, the channel supports both valley-chiral \cite{wu_valley-dependent_2011} and counterpropagating electron or hole modes \cite{pereira_confined_2006,beenakker_quantum_2009,yokoyama_gate-controlled_2010,biswas_scattering_2011}. We show the full phase diagram in Fig.\ \ref{fig:channel}(b) where each color-coded region corresponds to a distinct channel configuration, encompassing all possible cases. For real $\zeta$, we have $\tau \alpha k_y < 0$ and we always have chiral modes, while for imaginary $\zeta$ there exist both chiral ($v_+/v_- > 0$) and counterpropagating regimes with $\sgn(k_y) = \sgn \left[ \tau \alpha \cot(-i\zeta) \pm \gamma \csc(i\zeta) \right]$. In the latter chiral regime, we can have $v_+ = v_- = -\alpha / \gamma$ for $\gamma^2 = (m-1/2)^2 \pi^2 + \alpha^2$, giving a 1D Dirac cone. Remarkably, the velocity of one of the modes vanishes for $\gamma = \pm \tau \alpha \sech \zeta$, shown as solid lines in Fig.\ \ref{fig:channel}, yielding a 1D flatband. In the presence of electron-electron interactions, the 1D channel thus realizes a gate-tunable platform for Luttinger liquid physics where one can tune \textit{in situ} between a chiral Luttinger liquid and the Tomonaga model \cite{tomonaga_remarks_1950} with a gate-tunable Luttinger parameter.
Additionally, the strength of the strain barrier $\alpha$, is tunable by the height mismatch and separation of the electrodes, as well as the lattice orientation. Moreover, consecutive nanoslides, i.e.\ corrugations or terraces, can give rise to an array of coupled nanowires, similar to quantum-wire networks in marginally-twisted bilayer graphene \cite{de_beule_aharonov-bohm_2020,de_beule_network_2021,chen_correlated_2020,hsu_general_2023,wang_electrically_2024,chang_quasi-two-dimensional_2025} and periodically-buckled monolayer graphene \cite{de_beule_network_2023}.

\textcolor{NavyBlue}{\textit{Local density of states}} --- Another observable of the graphene nanoslide is the local density of states (LDOS) $\rho(x,E) = \rho_s(x,E) + \rho_b(x,E)$, which has contributions from scattering and bound states:
\begin{align}
    \rho_s(x,E) & = \sum_n \int_0^\infty \frac{dq}{2\pi} \int_{-\infty}^\infty \frac{dk_y}{2\pi} |\psi_n|^2 \delta(E - E_n), \label{eq:ldosS} \\
    \rho_b(x,E) & = \sum_n \int_{-\infty}^\infty \frac{dk_y}{2\pi} |\psi_n|^2 \delta(E - E_n).\label{eq:ldosB}
\end{align}
where $n$ runs over spin, valley, and different scattering or bound states.
From Eq.\ \eqref{eq:boundwave}, we find
\begin{equation}
    \rho_b(x, E) = \frac{2k_F}{\pi v_F} |\sinh \alpha| \cosh \alpha \, e^{-2|k_Fx \sinh \alpha|},
\end{equation}
which integrates to $2 / ( \pi v_F \sech \alpha )$ as expected for a 1D linear dispersion.
Next, we calculate the contribution from scattering states. Because of mirror symmetry ($x\mapsto-x,\sigma_y$) the LDOS is even in $x$ and we only have to consider $x < 0$. We obtain (details in SM \cite{Note1})
\begin{widetext}
\begin{equation}
    \frac{\delta \rho_s(x, E)}{\rho_0} = \frac{\rho_s(x, E) - \rho_0}{\rho_0} = - \left| \sinh \alpha \right| \left[ e^{-|\alpha|} J_0(2k_Fx) + 2 \cosh \alpha \sum_{n=1}^\infty J_{2n}(2k_Fx) e^{-2n |\alpha|} \right],
\end{equation}
\end{widetext}
with $\rho_0 = 2k_F / (\pi v_F)$ the bulk density of states and $J_{2n}(z)$ Bessel functions of the first kind. Thus, the scattering contribution decays as $|z|^{-1/2}$ while the bound state decays exponentially, see Fig.\ \ref{fig:ldos}(a). We further find $\rho(x=0,E) = \rho_0 \cosh^2 \alpha$ and $\rho(x,E) = \rho(x,-E)$ because of sublattice symmetry $\{ H, \sigma_z \} = 0$, which is broken by an electrostatic barrier  \cite{yokoyama_gate-controlled_2010}. We also consider the sublattice-resolved LDOS by letting $|\psi|^2 \rightarrow \psi^\dag \sigma_z \psi$ in Eqs.\ \eqref{eq:ldosS} and \eqref{eq:ldosB}. This gives an odd function of $x$ as the mirror symmetry exchanges sublattices, and can also be computed analytically \cite{Note1}. The pseudogauge barrier induces strong sublattice-dependent oscillations from the scattering contribution that are opposite on opposing sides of the barrier, consistent with symmetry, and shown in Fig. \ref{fig:ldos}(b). Moreover, $\rho_A$ and $\rho_B$ are out-of-phase for the two sublattices, leading to an approximate cancellation away from the barrier. The discontinuity at the origin stems from the boundary condition [Eq.\ \eqref{eq:boundary}].

We now turn on the electrostatic barrier from the bottom gate and compute the LDOS numerically. Note that for finite $\gamma = v_F^{-1} \int dx \, V(x)$, sublattice symmetry is broken, such that the LDOS is no longer electron-hole symmetric. Moreover, at resonance, we have perfect transmission and no bound states, such that fluctuations vanish. Thus, we expect that the LDOS can be strongly tuned with the bottom gate. In Fig.\ \ref{fig:ldos}(c), we show $\rho_A(x)$ from tight-binding simulations for fixed density in the leads,
versus $x$ and $V_b$ where we model the bottom gate with a potential $V(x)$ centered at $x=0$ with height $V_b$ and width $l$. The LDOS fluctuations for $V_b = 0$ are shown in Fig.\ \ref{fig:ldos}(d), showing good agreement between the lattice model and the analytical theory except near $x = 0$. This is expected because the only length scale for a delta barrier is $\lambda_F$ and smaller features cannot be resolved. Regardless, the sublattice Friedel oscillations and transport properties are in excellent agreement. 
For finite $V_b$, there is an electron-hole asymmetry that obeys $\rho_{A/B}(x, n, V_b) = \rho_{B/A}(-x,-n,-V_b)$.
In general, the electrostatic barrier suppresses the sublattice asymmetry, as shown in Fig.\ \ref{fig:ldos}(e).
\begin{figure}
    \includegraphics[width=\linewidth]{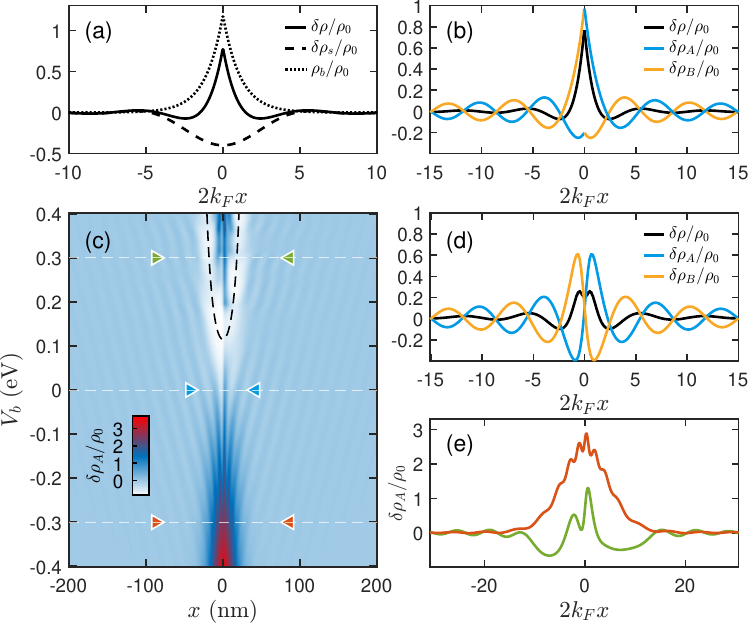}
    \caption{\textbf{Local density of states.} (a)--(b) Analytical result of the low-energy theory for $\alpha = 0.8$ and $\gamma = 0$. (a) Fluctuations of the LDOS (solid) showing contributions from scattering (dashed) and bound (dotted) states. (b) Sublattice-resolved LDOS fluctuations. (c)--(e) Tight-binding results for $\Delta = 1 \unit{nm}$, $d = 10 \unit{nm}$, $n = 10^{12} \unit{cm^{-2}}$, and potential $V(x) = V_b \sech^4(4x/l)$ with $l = 100 \unit{nm}$. (c) LDOS fluctuations for sublattice A versus $x$ and $V_b$. Triangles at $V_b = 0, \pm 0.3 \unit{eV}$ indicate the ranges of line cuts in (d) and (e).}
    \label{fig:ldos}
\end{figure}

\textcolor{NavyBlue}{\textit{Discussion}} --- In summary, we have theoretically investigated the electronic properties of the graphene nanoslide, which realizes a single strain-induced pseudogauge barrier. Using approximations that are physically justified and verified with extensive tight-binding simulations \cite{de_beule_preparation_2026}, we have obtained analytical solutions of the scattering problem, bound states, and local density of states. We have shown that both the longitudinal transport properties and transverse channel can be tuned \textit{in situ} by an additional electrostatic tunnel barrier using a bottom gate electrode. In particular, the nanoslide displays scattering resonances whenever a bound state merges with the Dirac continuum. These bound states form a one-dimensional channel that coexists with the continuum, with a rich phase diagram that supports both valley-chiral and counterpropagating modes. Our results are in qualitative agreement with recent experiments and establish the non-interacting electronic theory of the graphene nanoslide.

\let\oldaddcontentsline\addcontentsline 
\renewcommand{\addcontentsline}[3]{} 
\begin{acknowledgments}
\textcolor{NavyBlue}{\textit{Acknowledgments}} --- We thank Yu-Ting Hsiao for bringing Ref.\ \cite{zhang_gate-tunable_2022} to our attention, and Eugene J. Mele, Tse-Ming Chen, Chung-Ting Ke, and Hsin-You Wu for helpful discussions. C.D.B.\ was supported by the U.S.\ Department of Energy under Grant No.\ DE-FG02-84ER45118, and gratefully acknowledges QFort (NCKU) for hosting them during a research visit in which part of this work was completed. M.-H.\ L.\ acknowledges the National Science and Technology Council (NSTC) of Taiwan (under grant numbers 114-2112-M-006-029-MY3 and 112-2112-M-006-019-MY3) for financial support and the National Center for High-performance Computing (NCHC) for providing computational and storage resources. L.C. acknowledges support from Research Foundation-Flanders (FWO) research project No.\ G0A5921N and the EOS project ShapeME. We acknowledge financial support from the Methusalem funding of the University of Antwerp. 
\end{acknowledgments}

\bibliography{references}

@article{hsu_general_2023,
	title = {General scattering and electronic states in a quantum-wire network of moiré systems},
	volume = {108},
	url = {https://link.aps.org/doi/10.1103/PhysRevB.108.L121409},
	doi = {10.1103/PhysRevB.108.L121409},
	number = {12},
	urldate = {2026-03-11},
	journal = {Physical Review B},
	publisher = {American Physical Society},
	author = {Hsu, Chen-Hsuan and Loss, Daniel and Klinovaja, Jelena},
	month = sep,
	year = {2023},
	pages = {L121409},
}

@article{wang_electrically_2024,
	title = {Electrically tunable correlated domain wall network in twisted bilayer graphene},
	volume = {11},
	issn = {2053-1583},
	url = {https://doi.org/10.1088/2053-1583/ad3b11},
	doi = {10.1088/2053-1583/ad3b11},
	number = {3},
	urldate = {2026-03-11},
	journal = {2D Materials},
	publisher = {IOP Publishing},
	author = {Wang, Hao-Chien and Hsu, Chen-Hsuan},
	month = apr,
	year = {2024},
	pages = {035007},
}

@article{de_beule_network_2021,
	title = {Network model and four-terminal transport in minimally twisted bilayer graphene},
	volume = {104},
	url = {https://link.aps.org/doi/10.1103/PhysRevB.104.195410},
	doi = {10.1103/PhysRevB.104.195410},
	number = {19},
	urldate = {2026-03-11},
	journal = {Physical Review B},
	publisher = {American Physical Society},
	author = {De Beule, Christophe and Dominguez, Fernando and Recher, Patrik},
	month = nov,
	year = {2021},
	pages = {195410},
}

@article{de_beule_aharonov-bohm_2020,
	title = {Aharonov-{Bohm} {Oscillations} in {Minimally} {Twisted} {Bilayer} {Graphene}},
	volume = {125},
	url = {https://link.aps.org/doi/10.1103/PhysRevLett.125.096402},
	doi = {10.1103/PhysRevLett.125.096402},
	number = {9},
	urldate = {2026-03-11},
	journal = {Physical Review Letters},
	publisher = {American Physical Society},
	author = {De Beule, C. and Dominguez, F. and Recher, P.},
	month = aug,
	year = {2020},
	pages = {096402},
}

@article{chen_correlated_2020,
	title = {Correlated states of a triangular net of coupled quantum wires: {Implications} for the phase diagram of marginally twisted bilayer graphene},
	volume = {101},
	shorttitle = {Correlated states of a triangular net of coupled quantum wires},
	url = {https://link.aps.org/doi/10.1103/PhysRevB.101.165431},
	doi = {10.1103/PhysRevB.101.165431},
	number = {16},
	urldate = {2026-03-11},
	journal = {Physical Review B},
	publisher = {American Physical Society},
	author = {Chen, Chuan and Castro Neto, A. H. and Pereira, Vitor M.},
	month = apr,
	year = {2020},
	pages = {165431},
}

@article{chang_quasi-two-dimensional_2025,
	title = {Quasi-two-dimensional spin helix and magnon-induced singularity in twisted bilayer graphene},
	volume = {7},
	url = {https://link.aps.org/doi/10.1103/h5sd-v51h},
	doi = {10.1103/h5sd-v51h},
	number = {3},
	urldate = {2026-03-11},
	journal = {Physical Review Research},
	publisher = {American Physical Society},
	author = {Chang, Yung-Yeh and Saito, Kazuma and Hsu, Chen-Hsuan},
	month = sep,
	year = {2025},
	pages = {L032066},
}

@article{cazalilla_quantum_2014,
	title = {Quantum {Spin} {Hall} {Effect} in {Two}-{Dimensional} {Crystals} of {Transition}-{Metal} {Dichalcogenides}},
	volume = {113},
	url = {https://link.aps.org/doi/10.1103/PhysRevLett.113.077201},
	doi = {10.1103/PhysRevLett.113.077201},
	number = {7},
	urldate = {2024-08-12},
	journal = {Physical Review Letters},
	author = {Cazalilla, M. A. and Ochoa, H. and Guinea, F.},
	month = aug,
	year = {2014},
	pages = {077201},
}

@article{castro_neto_electronic_2009,
	title = {The electronic properties of graphene},
	volume = {81},
	url = {https://link.aps.org/doi/10.1103/RevModPhys.81.109},
	doi = {10.1103/RevModPhys.81.109},
	number = {1},
	urldate = {2023-09-24},
	journal = {Reviews of Modern Physics},
	author = {Castro Neto, A. H. and Guinea, F. and Peres, N. M. R. and Novoselov, K. S. and Geim, A. K.},
	month = jan,
	year = {2009},
	pages = {109--162},
}

@article{zhang_gate-tunable_2022,
	title = {Gate-tunable {Veselago} interference in a bipolar graphene microcavity},
	volume = {13},
	copyright = {2022 The Author(s)},
	issn = {2041-1723},
	url = {https://www.nature.com/articles/s41467-022-34347-w},
	doi = {10.1038/s41467-022-34347-w},
	number = {1},
	urldate = {2025-02-03},
	journal = {Nature Communications},
	author = {Zhang, Xi and Ren, Wei and Bell, Elliot and Zhu, Ziyan and Tsai, Kan-Ting and Luo, Yujie and Watanabe, Kenji and Taniguchi, Takashi and Kaxiras, Efthimios and Luskin, Mitchell and Wang, Ke},
	month = nov,
	year = {2022},
	pages = {6711},
}

@article{wehling_midgap_2008,
	title = {Midgap states in corrugated graphene: {Ab} initio calculations and effective field theory},
	volume = {84},
	issn = {0295-5075},
	shorttitle = {Midgap states in corrugated graphene},
	url = {https://dx.doi.org/10.1209/0295-5075/84/17003},
	doi = {10.1209/0295-5075/84/17003},
	number = {1},
	urldate = {2024-02-08},
	journal = {Europhysics Letters},
	author = {Wehling, T. O. and Balatsky, A. V. and Tsvelik, A. M. and Katsnelson, M. I. and Lichtenstein, A. I.},
	month = sep,
	year = {2008},
	pages = {17003},
}

@article{wang_valley-polarized_2023,
	title = {Valley-polarized and enhanced transmission in graphene with a smooth strain profile},
	volume = {35},
	issn = {0953-8984},
	url = {https://dx.doi.org/10.1088/1361-648X/accbf9},
	doi = {10.1088/1361-648X/accbf9},
	number = {30},
	urldate = {2025-02-03},
	journal = {Journal of Physics: Condensed Matter},
	author = {Wang, Sake and Tian, Hongyu and Sun, Minglei},
	month = apr,
	year = {2023},
	pages = {304002},
}

@article{suzuura_phonons_2002,
	title = {Phonons and electron-phonon scattering in carbon nanotubes},
	volume = {65},
	url = {https://link.aps.org/doi/10.1103/PhysRevB.65.235412},
	doi = {10.1103/PhysRevB.65.235412},
	number = {23},
	urldate = {2023-09-24},
	journal = {Physical Review B},
	author = {Suzuura, Hidekatsu and Ando, Tsuneya},
	month = may,
	year = {2002},
	pages = {235412},
}

@article{rostami_theory_2015,
	title = {Theory of strain in single-layer transition metal dichalcogenides},
	volume = {92},
	url = {https://link.aps.org/doi/10.1103/PhysRevB.92.195402},
	doi = {10.1103/PhysRevB.92.195402},
	number = {19},
	urldate = {2024-08-22},
	journal = {Physical Review B},
	author = {Rostami, Habib and Roldán, Rafael and Cappelluti, Emmanuele and Asgari, Reza and Guinea, Francisco},
	month = nov,
	year = {2015},
	pages = {195402},
}

@article{roldan_strain_2015,
	title = {Strain engineering in semiconducting two-dimensional crystals},
	volume = {27},
	issn = {0953-8984},
	url = {https://dx.doi.org/10.1088/0953-8984/27/31/313201},
	doi = {10.1088/0953-8984/27/31/313201},
	number = {31},
	urldate = {2024-08-12},
	journal = {Journal of Physics: Condensed Matter},
	author = {Roldán, Rafael and Castellanos-Gomez, Andrés and Cappelluti, Emmanuele and Guinea, Francisco},
	month = jul,
	year = {2015},
	pages = {313201},
}

@article{phong_boundary_2022,
	title = {Boundary {Modes} from {Periodic} {Magnetic} and {Pseudomagnetic} {Fields} in {Graphene}},
	volume = {128},
	url = {https://link.aps.org/doi/10.1103/PhysRevLett.128.176406},
	doi = {10.1103/PhysRevLett.128.176406},
	number = {17},
	urldate = {2024-07-31},
	journal = {Physical Review Letters},
	author = {Phong, Võ Tiến and Mele, E. J.},
	month = apr,
	year = {2022},
	pages = {176406},
}

@article{pikulin_chiral_2016,
	title = {Chiral {Anomaly} from {Strain}-{Induced} {Gauge} {Fields} in {Dirac} and {Weyl} {Semimetals}},
	volume = {6},
	url = {https://link.aps.org/doi/10.1103/PhysRevX.6.041021},
	doi = {10.1103/PhysRevX.6.041021},
	number = {4},
	urldate = {2024-08-12},
	journal = {Physical Review X},
	author = {Pikulin, D. I. and Chen, Anffany and Franz, M.},
	month = oct,
	year = {2016},
	pages = {041021},
}

@article{pereira_tight-binding_2009,
	title = {Tight-binding approach to uniaxial strain in graphene},
	volume = {80},
	url = {https://link.aps.org/doi/10.1103/PhysRevB.80.045401},
	doi = {10.1103/PhysRevB.80.045401},
	number = {4},
	urldate = {2025-02-08},
	journal = {Physical Review B},
	author = {Pereira, Vitor M. and Castro Neto, A. H. and Peres, N. M. R.},
	month = jul,
	year = {2009},
	pages = {045401},
}

@article{nigge_room_2019,
	title = {Room temperature strain-induced {Landau} levels in graphene on a wafer-scale platform},
	volume = {5},
	url = {https://www.science.org/doi/10.1126/sciadv.aaw5593},
	doi = {10.1126/sciadv.aaw5593},
	number = {11},
	urldate = {2024-08-12},
	journal = {Science Advances},
	author = {Nigge, P. and Qu, A. C. and Lantagne-Hurtubise, E. and Marsell, E. and Link, S. and Tom, G. and Zonno, M. and Michiardi, M. and Schneider, M. and Zhdanovich, S. and Levy, G. and Starke, U. and Guti{\textbackslash}'errez, C. and Bonn, D. and Burke, S. A. and Franz, M. and Damascelli, A.},
	month = nov,
	year = {2019},
	pages = {eaaw5593},
}

@article{milovanovic_band_2020,
	title = {Band flattening in buckled monolayer graphene},
	volume = {102},
	url = {https://link.aps.org/doi/10.1103/PhysRevB.102.245427},
	doi = {10.1103/PhysRevB.102.245427},
	number = {24},
	urldate = {2024-07-31},
	journal = {Physical Review B},
	author = {Milovanović, S. P. and Anđelković, M. and Covaci, L. and Peeters, F. M.},
	month = dec,
	year = {2020},
	pages = {245427},
}

@article{mao_evidence_2020,
	title = {Evidence of flat bands and correlated states in buckled graphene superlattices},
	volume = {584},
	copyright = {2020 The Author(s), under exclusive licence to Springer Nature Limited},
	issn = {1476-4687},
	url = {https://www.nature.com/articles/s41586-020-2567-3},
	doi = {10.1038/s41586-020-2567-3},
	number = {7820},
	urldate = {2023-09-24},
	journal = {Nature},
	author = {Mao, Jinhai and Milovanović, Slaviša P. and Anđelković, Miša and Lai, Xinyuan and Cao, Yang and Watanabe, Kenji and Taniguchi, Takashi and Covaci, Lucian and Peeters, Francois M. and Geim, Andre K. and Jiang, Yuhang and Andrei, Eva Y.},
	month = aug,
	year = {2020},
	pages = {215--220},
}

@article{manesco_correlation-induced_2021,
	title = {Correlation-induced valley topology in buckled graphene superlattices},
	volume = {8},
	issn = {2053-1583},
	url = {https://dx.doi.org/10.1088/2053-1583/ac0b48},
	doi = {10.1088/2053-1583/ac0b48},
	number = {3},
	urldate = {2023-09-20},
	journal = {2D Materials},
	author = {Manesco, Antonio L. R. and Lado, Jose L.},
	month = jun,
	year = {2021},
	pages = {035057},
}

@article{manesco_correlations_2020,
	title = {Correlations in the elastic {Landau} level of spontaneously buckled graphene},
	volume = {8},
	issn = {2053-1583},
	url = {https://dx.doi.org/10.1088/2053-1583/abbc5f},
	doi = {10.1088/2053-1583/abbc5f},
	number = {1},
	urldate = {2023-09-20},
	journal = {2D Materials},
	author = {Manesco, A. L. R. and Lado, J. L. and Ribeiro, E. V. S. and Weber, G. and Jr, D. Rodrigues},
	month = oct,
	year = {2020},
	pages = {015011},
}

@article{levy_strain-induced_2010,
	title = {Strain-{Induced} {Pseudo}–{Magnetic} {Fields} {Greater} {Than} 300 {Tesla} in {Graphene} {Nanobubbles}},
	volume = {329},
	url = {https://www.science.org/doi/10.1126/science.1191700},
	doi = {10.1126/science.1191700},
	number = {5991},
	urldate = {2024-08-12},
	journal = {Science},
	author = {Levy, N. and Burke, S. A. and Meaker, K. L. and Panlasigui, M. and Zettl, A. and Guinea, F. and Neto, A. H. Castro and Crommie, M. F.},
	month = jul,
	year = {2010},
	pages = {544--547},
}

@article{katsnelson_graphene_2007,
	series = {Exploring graphene},
	title = {Graphene: {New} bridge between condensed matter physics and quantum electrodynamics},
	volume = {143},
	issn = {0038-1098},
	shorttitle = {Graphene},
	url = {https://www.sciencedirect.com/science/article/pii/S0038109807003043},
	doi = {10.1016/j.ssc.2007.02.043},
	number = {1},
	urldate = {2024-02-12},
	journal = {Solid State Communications},
	author = {Katsnelson, M. I. and Novoselov, K. S.},
	month = jul,
	year = {2007},
	pages = {3--13},
}

@article{kane_size_1997,
	title = {Size, {Shape}, and {Low} {Energy} {Electronic} {Structure} of {Carbon} {Nanotubes}},
	volume = {78},
	url = {https://link.aps.org/doi/10.1103/PhysRevLett.78.1932},
	doi = {10.1103/PhysRevLett.78.1932},
	number = {10},
	urldate = {2024-07-22},
	journal = {Physical Review Letters},
	author = {Kane, C. L. and Mele, E. J.},
	month = mar,
	year = {1997},
	pages = {1932--1935},
}

@article{iordanskii_dislocations_1985,
	title = {Dislocations and localization effects in multivalley conductors},
	volume = {41},
	url = {http://jetpletters.ru/ps/0/article_22404.shtml},
	number = {11},
	journal = {JETP Lett.},
	author = {Iordanskii, S. V and Koshelev, A. E.},
	year = {1985},
	pages = {471},
}

@article{ilan_pseudo-electromagnetic_2020,
	title = {Pseudo-electromagnetic fields in {3D} topological semimetals},
	volume = {2},
	copyright = {2019 Springer Nature Limited},
	issn = {2522-5820},
	url = {https://www.nature.com/articles/s42254-019-0121-8},
	doi = {10.1038/s42254-019-0121-8},
	number = {1},
	urldate = {2024-08-12},
	journal = {Nature Reviews Physics},
	author = {Ilan, Roni and Grushin, Adolfo G. and Pikulin, Dmitry I.},
	month = jan,
	year = {2020},
	pages = {29--41},
}

@article{ho_hall_2021,
	title = {Hall effects in artificially corrugated bilayer graphene without breaking time-reversal symmetry},
	volume = {4},
	copyright = {2021 The Author(s), under exclusive licence to Springer Nature Limited},
	issn = {2520-1131},
	url = {https://www.nature.com/articles/s41928-021-00537-5},
	doi = {10.1038/s41928-021-00537-5},
	number = {2},
	urldate = {2023-09-28},
	journal = {Nature Electronics},
	author = {Ho, Sheng-Chin and Chang, Ching-Hao and Hsieh, Yu-Chiang and Lo, Shun-Tsung and Huang, Botsz and Vu, Thi-Hai-Yen and Ortix, Carmine and Chen, Tse-Ming},
	month = feb,
	year = {2021},
	pages = {116--125},
}

@article{guinea_energy_2010,
	title = {Energy gaps and a zero-field quantum {Hall} effect in graphene by strain engineering},
	volume = {6},
	copyright = {2009 Springer Nature Limited},
	issn = {1745-2481},
	url = {https://www.nature.com/articles/nphys1420},
	doi = {10.1038/nphys1420},
	number = {1},
	urldate = {2024-08-12},
	journal = {Nature Physics},
	author = {Guinea, F. and Katsnelson, M. I. and Geim, A. K.},
	month = jan,
	year = {2010},
	pages = {30--33},
}

@article{guinea_gauge_2008,
	title = {Gauge field induced by ripples in graphene},
	volume = {77},
	url = {https://link.aps.org/doi/10.1103/PhysRevB.77.205421},
	doi = {10.1103/PhysRevB.77.205421},
	number = {20},
	urldate = {2023-09-24},
	journal = {Physical Review B},
	author = {Guinea, F. and Horovitz, Baruch and Le Doussal, P.},
	month = may,
	year = {2008},
	pages = {205421},
}

@article{guinea_generating_2010,
	title = {Generating quantizing pseudomagnetic fields by bending graphene ribbons},
	volume = {81},
	url = {https://link.aps.org/doi/10.1103/PhysRevB.81.035408},
	doi = {10.1103/PhysRevB.81.035408},
	number = {3},
	urldate = {2025-02-08},
	journal = {Physical Review B},
	author = {Guinea, F. and Geim, A. K. and Katsnelson, M. I. and Novoselov, K. S.},
	month = jan,
	year = {2010},
	pages = {035408},
}

@article{gao_untwisting_2023,
	title = {Untwisting {Moiré} {Physics}: {Almost} {Ideal} {Bands} and {Fractional} {Chern} {Insulators} in {Periodically} {Strained} {Monolayer} {Graphene}},
	volume = {131},
	shorttitle = {Untwisting {Moir}{\textbackslash}'e {Physics}},
	url = {https://link.aps.org/doi/10.1103/PhysRevLett.131.096401},
	doi = {10.1103/PhysRevLett.131.096401},
	number = {9},
	urldate = {2023-09-20},
	journal = {Physical Review Letters},
	author = {Gao, Qiang and Dong, Junkai and Ledwith, Patrick and Parker, Daniel and Khalaf, Eslam},
	month = aug,
	year = {2023},
	pages = {096401},
}

@article{de_beule_network_2023,
	title = {Network model for periodically strained graphene},
	volume = {107},
	url = {https://link.aps.org/doi/10.1103/PhysRevB.107.045405},
	doi = {10.1103/PhysRevB.107.045405},
	number = {4},
	urldate = {2025-03-27},
	journal = {Physical Review B},
	author = {De Beule, Christophe and Phong, Võ Tiến and Mele, E. J.},
	month = jan,
	year = {2023},
	pages = {045405},
}

@article{cortijo_elastic_2015,
	title = {Elastic {Gauge} {Fields} in {Weyl} {Semimetals}},
	volume = {115},
	url = {https://link.aps.org/doi/10.1103/PhysRevLett.115.177202},
	doi = {10.1103/PhysRevLett.115.177202},
	number = {17},
	urldate = {2024-08-12},
	journal = {Physical Review Letters},
	author = {Cortijo, Alberto and Ferreirós, Yago and Landsteiner, Karl and Vozmediano, María A. H.},
	month = oct,
	year = {2015},
	pages = {177202},
}

@article{yokoyama_gate-controlled_2010,
	title = {Gate-{Controlled} {One}-{Dimensional} {Channel} on the {Surface} of a {3D} {Topological} {Insulator}},
	volume = {104},
	url = {https://link.aps.org/doi/10.1103/PhysRevLett.104.246806},
	doi = {10.1103/PhysRevLett.104.246806},
	number = {24},
	urldate = {2025-12-03},
	journal = {Physical Review Letters},
	author = {Yokoyama, Takehito and Balatsky, Alexander V. and Nagaosa, Naoto},
	month = jun,
	year = {2010},
	pages = {246806},
}

@article{wu_valley-dependent_2011,
	title = {Valley-{Dependent} {Brewster} {Angles} and {Goos}-{Hänchen} {Effect} in {Strained} {Graphene}},
	volume = {106},
	url = {https://link.aps.org/doi/10.1103/PhysRevLett.106.176802},
	doi = {10.1103/PhysRevLett.106.176802},
	number = {17},
	urldate = {2025-12-12},
	journal = {Physical Review Letters},
	author = {Wu, Zhenhua and Zhai, F. and Peeters, F. M. and Xu, H. Q. and Chang, Kai},
	month = apr,
	year = {2011},
	pages = {176802},
}

@article{wu_quantum_2018,
	title = {Quantum {Wires} and {Waveguides} {Formed} in {Graphene} by {Strain}},
	volume = {18},
	issn = {1530-6984},
	url = {https://doi.org/10.1021/acs.nanolett.7b03167},
	doi = {10.1021/acs.nanolett.7b03167},
	number = {1},
	urldate = {2025-12-11},
	journal = {Nano Letters},
	author = {Wu, Y. and Zhai, D. and Pan, C. and Cheng, B. and Taniguchi, T. and Watanabe, K. and Sandler, N. and Bockrath, M.},
	month = jan,
	year = {2018},
	pages = {64--69},
}

@article{venderbos_interacting_2016,
	title = {Interacting {Dirac} fermions under a spatially alternating pseudomagnetic field: {Realization} of spontaneous quantum {Hall} effect},
	volume = {93},
	shorttitle = {Interacting {Dirac} fermions under a spatially alternating pseudomagnetic field},
	url = {https://link.aps.org/doi/10.1103/PhysRevB.93.195126},
	doi = {10.1103/PhysRevB.93.195126},
	number = {19},
	urldate = {2026-02-11},
	journal = {Physical Review B},
	publisher = {American Physical Society},
	author = {Venderbos, Jörn W. F. and Fu, Liang},
	month = may,
	year = {2016},
	pages = {195126},
}

@article{tomonaga_remarks_1950,
	title = {Remarks on {Bloch}'s {Method} of {Sound} {Waves} applied to {Many}-{Fermion} {Problems}},
	volume = {5},
	issn = {0033-068X},
	url = {https://doi.org/10.1143/ptp/5.4.544},
	doi = {10.1143/ptp/5.4.544},
	number = {4},
	urldate = {2025-12-04},
	journal = {Progress of Theoretical Physics},
	author = {Tomonaga, Sin-Itiro},
	month = jul,
	year = {1950},
	pages = {544--569},
}

@article{timm_transport_2012,
	title = {Transport through a quantum spin {Hall} quantum dot},
	volume = {86},
	url = {https://link.aps.org/doi/10.1103/PhysRevB.86.155456},
	doi = {10.1103/PhysRevB.86.155456},
	number = {15},
	urldate = {2025-12-04},
	journal = {Physical Review B},
	author = {Timm, Carsten},
	month = oct,
	year = {2012},
	pages = {155456},
}

@article{pereira_confined_2006,
	title = {Confined states and direction-dependent transmission in graphene quantum wells},
	volume = {74},
	url = {https://link.aps.org/doi/10.1103/PhysRevB.74.045424},
	doi = {10.1103/PhysRevB.74.045424},
	number = {4},
	urldate = {2025-12-26},
	journal = {Physical Review B},
	publisher = {American Physical Society},
	author = {Pereira, J. Milton and Mlinar, V. and Peeters, F. M. and Vasilopoulos, P.},
	month = jul,
	year = {2006},
	pages = {045424},
}

@article{vozmediano_gauge_2010,
	title = {Gauge fields in graphene},
	volume = {496},
	issn = {0370-1573},
	url = {https://www.sciencedirect.com/science/article/pii/S0370157310001729},
	doi = {10.1016/j.physrep.2010.07.003},
	number = {4},
	urldate = {2023-09-24},
	journal = {Physics Reports},
	author = {Vozmediano, M. A. H. and Katsnelson, M. I. and Guinea, F.},
	month = nov,
	year = {2010},
	pages = {109--148},
}

@article{nakatsuji_uniaxial_2012,
	title = {Uniaxial deformation of graphene {Dirac} cone on a vicinal {SiC} substrate},
	volume = {85},
	url = {https://link.aps.org/doi/10.1103/PhysRevB.85.195416},
	doi = {10.1103/PhysRevB.85.195416},
	number = {19},
	urldate = {2025-12-24},
	journal = {Physical Review B},
	author = {Nakatsuji, Kan and Yoshimura, Tsuguo and Komori, Fumio and Morita, Kouhei and Tanaka, Satoru},
	month = may,
	year = {2012},
	pages = {195416},
}

@article{low_deformation_2012,
	title = {Deformation and {Scattering} in {Graphene} over {Substrate} {Steps}},
	volume = {108},
	url = {https://link.aps.org/doi/10.1103/PhysRevLett.108.096601},
	doi = {10.1103/PhysRevLett.108.096601},
	number = {9},
	urldate = {2025-12-24},
	journal = {Physical Review Letters},
	author = {Low, T. and Perebeinos, V. and Tersoff, J. and Avouris, Ph.},
	month = mar,
	year = {2012},
	pages = {096601},
}

@article{liu_snake_2015,
	title = {Snake states and their symmetries in graphene},
	volume = {92},
	url = {https://link.aps.org/doi/10.1103/PhysRevB.92.235438},
	doi = {10.1103/PhysRevB.92.235438},
	number = {23},
	urldate = {2025-12-23},
	journal = {Physical Review B},
	author = {Liu, Yang and Tiwari, Rakesh P. and Brada, Matej and Bruder, C. and Kusmartsev, F. V. and Mele, E. J.},
	month = dec,
	year = {2015},
	pages = {235438},
}

@article{jun_nanowrinkle_2025,
	title = {Nanowrinkle waveguide in graphene for enabling secure {Dirac} fermion transport},
	volume = {58},
	issn = {0022-3727},
	url = {https://doi.org/10.1088/1361-6463/ada44e},
	doi = {10.1088/1361-6463/ada44e},
	number = {11},
	urldate = {2025-12-11},
	journal = {Journal of Physics D: Applied Physics},
	author = {Jun, Seunghyun and Jung, Myung-Chul and Myoung, Nojoon},
	month = jan,
	year = {2025},
	pages = {115301},
}

@article{divincenzo_self-consistent_1984,
	title = {Self-consistent effective-mass theory for intralayer screening in graphite intercalation compounds},
	volume = {29},
	url = {https://link.aps.org/doi/10.1103/PhysRevB.29.1685},
	doi = {10.1103/PhysRevB.29.1685},
	number = {4},
	urldate = {2025-12-05},
	journal = {Physical Review B},
	author = {DiVincenzo, D. P. and Mele, E. J.},
	month = feb,
	year = {1984},
	pages = {1685--1694},
}

@misc{de_beule_preparation_2026,
	title = {In preparation},
	author = {De Beule, Christophe and Wu, Hsin-You and Garcia-Ruiz, Aitor and Kang, Wun-Hao and Mrénca-Kolasínska, Alina and Liu, Ming-Hao},
	year = {2026},
}

@article{chakraborti_electron_2024,
	title = {Electron wave and quantum optics in graphene},
	volume = {36},
	issn = {0953-8984},
	url = {https://doi.org/10.1088/1361-648X/ad46bc},
	doi = {10.1088/1361-648X/ad46bc},
	number = {39},
	urldate = {2025-12-21},
	journal = {Journal of Physics: Condensed Matter},
	author = {Chakraborti, Himadri and Gorini, Cosimo and Knothe, Angelika and Liu, Ming-Hao and Makk, Péter and Parmentier, François D and Perconte, David and Richter, Klaus and Roulleau, Preden and Sacépé, Benjamin and Schönenberger, Christian and Yang, Wenmin},
	month = jul,
	year = {2024},
	pages = {393001},
}

@article{beenakker_quantum_2009,
	title = {Quantum {Goos}-{Hänchen} {Effect} in {Graphene}},
	volume = {102},
	url = {https://link.aps.org/doi/10.1103/PhysRevLett.102.146804},
	doi = {10.1103/PhysRevLett.102.146804},
	number = {14},
	urldate = {2025-12-26},
	journal = {Physical Review Letters},
	author = {Beenakker, C. W. J. and Sepkhanov, R. A. and Akhmerov, A. R. and Tworzydło, J.},
	month = apr,
	year = {2009},
	pages = {146804},
}

@article{biswas_scattering_2011,
	title = {Scattering from surface step edges in strong topological insulators},
	volume = {83},
	url = {https://link.aps.org/doi/10.1103/PhysRevB.83.075439},
	doi = {10.1103/PhysRevB.83.075439},
	number = {7},
	urldate = {2025-12-03},
	journal = {Physical Review B},
	author = {Biswas, Rudro R. and Balatsky, Alexander V.},
	month = feb,
	year = {2011},
	pages = {075439},
}

@article{baringhaus_exceptional_2014,
	title = {Exceptional ballistic transport in epitaxial graphene nanoribbons},
	volume = {506},
	copyright = {2014 Springer Nature Limited},
	issn = {1476-4687},
	url = {https://www.nature.com/articles/nature12952},
	doi = {10.1038/nature12952},
	number = {7488},
	urldate = {2026-01-09},
	journal = {Nature},
	author = {Baringhaus, Jens and Ruan, Ming and Edler, Frederik and Tejeda, Antonio and Sicot, Muriel and Taleb-Ibrahimi, Amina and Li, An-Ping and Jiang, Zhigang and Conrad, Edward H. and Berger, Claire and Tegenkamp, Christoph and de Heer, Walt A.},
	month = feb,
	year = {2014},
	pages = {349--354},
}

@article{banerjee_strain_2020,
	title = {Strain {Modulated} {Superlattices} in {Graphene}},
	volume = {20},
	issn = {1530-6984},
	url = {https://doi.org/10.1021/acs.nanolett.9b05108},
	doi = {10.1021/acs.nanolett.9b05108},
	number = {5},
	urldate = {2025-12-05},
	journal = {Nano Letters},
	author = {Banerjee, Riju and Nguyen, Viet-Hung and Granzier-Nakajima, Tomotaroh and Pabbi, Lavish and Lherbier, Aurelien and Binion, Anna Ruth and Charlier, Jean-Christophe and Terrones, Mauricio and Hudson, Eric William},
	month = may,
	year = {2020},
	pages = {3113--3121},
}

@article{low_gaps_2011,
	title = {Gaps tunable by electrostatic gates in strained graphene},
	volume = {83},
	url = {https://link.aps.org/doi/10.1103/PhysRevB.83.195436},
	doi = {10.1103/PhysRevB.83.195436},
	number = {19},
	urldate = {2023-09-23},
	journal = {Physical Review B},
	author = {Low, T. and Guinea, F. and Katsnelson, M. I.},
	month = may,
	year = {2011},
	pages = {195436},
}
\let\addcontentsline\oldaddcontentsline 

\clearpage
\begin{CJK}{UTF8}{bsmi}

\title{Supplemental Material for: \mytitle}

\author{Christophe De Beule}
\email{christophe.debeule@uantwerpen.be}
\affiliation{Department of Physics and Astronomy, University of Pennsylvania, Philadelphia, Pennsylvania 19104, USA}
\affiliation{Department of Physics, University of Antwerp, Groenenborgerlaan 171, 2020 Antwerp, Belgium}

\author{Ming-Hao Liu (劉明豪)}
\email{minghao.liu@phys.ncku.edu.tw}
\affiliation{Department of Physics and Center for Quantum Frontiers of Research and Technology (QFort), National Cheng Kung University, Tainan 70101, Taiwan}

\author{Bart Partoens}
\affiliation{Department of Physics, University of Antwerp, Groenenborgerlaan 171, 2020 Antwerp, Belgium}

\author{Lucian Covaci}
\affiliation{Department of Physics and NANOlight Center of Excellence, University of Antwerp, Groenenborgerlaan 171, 2020 Antwerp, Belgium}

\maketitle
\end{CJK}

\onecolumngrid
 
\setcounter{equation}{0}
\setcounter{figure}{0}
\setcounter{table}{0}
\setcounter{page}{1}
\setcounter{secnumdepth}{2}
\makeatletter
\renewcommand{\thepage}{S\arabic{page}}
\renewcommand{\thesection}{S\arabic{section}}
\renewcommand{\theequation}{S\arabic{equation}}
\renewcommand{\thefigure}{S\arabic{figure}}
\renewcommand{\thetable}{S\arabic{table}}

\tableofcontents
\vspace{1cm}

\twocolumngrid

\section{In-plane lattice relaxation}

\subsection{Relaxed uniaxial strain}

In the presence of elastic deformations that only depend on the $x$ direction, the structural energy of graphene can be modeled as \cite{guinea_gauge_2008}
\begin{equation}
    H_\text{elas} + H_\text{subs} = \int dx \, \mathcal F \left( \partial_x \bm u; h, \partial_x h, \partial_x^2 h \right),
\end{equation}
where $\bm u = (u_x, u_y)$ is the in-plane displacement field and $h$ is the out-of-plane displacement field. We further have
\begin{align}
    \mathcal F_\text{elas} & = \frac{1}{2} \left( \lambda + 2 \mu \right) u_{xx}^2 + 2 \mu u_{xy}^2 + \frac{\kappa}{2} \left( \frac{d^2h}{dx^2} \right)^2,\\
    \mathcal F_\text{subs} & = \frac{g}{2} \left[ h(x) - s(x) - h_\text{eq} \right]^2,
\end{align}
where $\lambda$ and $\mu$ are in-plane Lam\'e parameters, $\kappa$ is the bending rigidity, $g$ is the coupling strength with the substrate $s(x)$, and $h_\text{eq}$ is the equilibrium distance. The equations of motions are
\begin{align}
    \frac{d}{dx} \frac{\partial \mathcal F}{\partial( \partial_x u_i )} & = 0, \\
    \frac{\partial \mathcal F}{\partial h} - \frac{d}{dx} \frac{\partial \mathcal F}{\partial( \partial_x h)} + \frac{d^2}{dx^2} \frac{\partial \mathcal F}{\partial( \partial_x^2 h)} & = 0.
\end{align}
This yields
\begin{equation}
    u_{xx} = c_1, \qquad u_{xy} = c_2,
\end{equation}
where $c_1$ and $c_2$ are constants, and
\begin{equation}
     c_1 \left( \lambda + 2\mu \right) \frac{d^2h}{dx^2} = \kappa \frac{d^4h}{dx^4} + g\left(h - s - h_\text{eq} \right).
\end{equation}

Therefore, this theory predicts that the pseudogauge field for relaxed uniaxial strain is constant and the pseudomagnetic field vanishes. This result was discussed in Ref.\ \cite{wehling_midgap_2008} and agrees with fully-relaxed microscopic density-functional theory calculations for a periodic corrugation where the maxima and minima were fixed.

Let us now consider a strain profile that varies on a length scale $L$. Then when $g L^4 \gg \kappa$ and $g L^2 \gg c_1 (\lambda + 2\mu)$, we have $h = h_\text{eq} + s$ and
\begin{equation}
    u_x(x) = c_1 x - \frac{1}{2} \int_{x_0}^x dx' \left[ \frac{ds(x')}{dx'} \right]^2, \quad u_y = 0.
\end{equation}
For example, consider a periodic corrugation \cite{wehling_midgap_2008, banerjee_strain_2020}
\begin{equation}
    s(x) = 2 h_0 \sin \left( \frac{2\pi x}{L} \right),
\end{equation}
such that $c_1 = (2\pi h_0/L)^2$, thus also requiring $\gamma L^4 \gg  4\pi^2 h_0^2 (\lambda + 2\mu)$. We then obtain
\begin{equation}
    u_x(x) = - \frac{\pi h_0^2 \sin(4\pi x/L)}{L}.
\end{equation}
where we disregarded any constant displacements. And indeed, we find
\begin{align}
    u_{xx} & = -\frac{4 \pi^2 h_0^2}{L^2} \, \cos \left( \frac{4\pi x}{L} \right) + \frac{1}{2} \left[ \frac{4\pi h_0}{L} \cos \left( \frac{2\pi x}{L} \right) \right]^2 \\
    & = \frac{4 \pi^2 h_0^2}{L^2} = c_1, \\
    u_{xy} & = 0.
\end{align}
For the graphene nanoslide, we have
\begin{equation}
    s(x) = \frac{\Delta}{2} \tanh \left( \frac{4x}{d} \right),
\end{equation}
and we require $c_1 = 0$. This gives
\begin{equation}
    u_x(x) = \frac{\Delta^2}{6d} \tanh \left( \frac{4x}{d} \right) \left[ \tanh^2 \left( \frac{4x}{d} \right) - 3 \right],
\end{equation}
such that $u_{xx} = u_{yy} = 0$.

\subsection{Phenomenological approach}

The theory in the previous section does not account for finite-size effects and more complicated substrate interactions. In order to model experiments, we introduce a phenomenological parameter $\lambda \in [0, 1]$ and replace $u_x(x) \rightarrow \lambda u_x(x)$. Here $\lambda = 0$ corresponds to a quenched corrugation (no in-plane relaxation) and $\lambda = 1$ corresponds to an annealed corrugation (fully relaxed). For the last case, we find that
\begin{equation}
    u_{xx}(x) \rightarrow \frac{2\Delta^2 ( 1 - \lambda )}{d^2} \, \mathrm{sech}^4 \left( \frac{4x}{d} \right),
\end{equation}
such that effectively we could also use $h \rightarrow h \sqrt{1 - \lambda}$ instead of the in-plane displacement field, as this results in the same strain tensor. This is true for a general height profile, because
\begin{align}
    u_{xx}(x) & \rightarrow \lambda \left[ c_1 - \frac{1}{2} \left( \frac{dh}{dx} \right)^2 \right] + \frac{1}{2} \left( \frac{dh}{dx} \right)^2 \\
    & = \lambda c_1 + \frac{1}{2} \left( \frac{dh}{dx} \sqrt{1 - \lambda} \right)^2,
\end{align}
where the constant $\lambda c_1$ only amounts to a trivial gauge transformation of the resulting pseudogauge field. This justifies our phenomenological approach to account for in-plane lattice relaxation with a reduction factor.

\section{Boundary conditions}

Here we present an alternative derivation of the boundary conditions for the pseudogauge delta barrier. We start by writing down the wave equation for the Dirac spinor $\Psi(x,y) = \psi(x) e^{ik_y y}$ with $\psi = (\psi_A, \psi_B)^\top$ where $\psi_A$ and $\psi_B$ are the sublattice components. We have
\begin{widetext}
\begin{align}
    -i \tau \frac{d}{dx} \psi_B - i \left[ k_y + \alpha \delta(x) \right] \psi_B & = \frac{E}{v_F} \psi_A, \\
    -i \tau \frac{d}{dx} \psi_A + i \left[ k_y + \alpha \delta(x) \right] \psi_A & = \frac{E}{v_F} \psi_B.
\end{align}
Note that we cannot integrate this equation directly around a small region centered at $x = 0$ because the spinor is not continuous and thus $\psi(0)$ is ill-defined. To avoid this, we multiply the first (second) equation by $\psi_A$ ($\psi_B$) and then add and subtract:
\begin{align}
    -i \tau \frac{d}{dx} \left( \psi_A \psi_B \right) = -i \tau \psi_A \frac{d}{dx} \psi_B - i \tau \psi_B \frac{d}{dx} \psi_A & = \frac{E}{v_F} \left( \psi_A^2 + \psi_B^2 \right), \\
    -i \tau \psi_A \frac{d}{dx} \psi_B + i \tau \psi_B \frac{d}{dx} \psi_A & = 2 \left[ k_y + \alpha \delta(x) \right] \psi_A \psi_B + \frac{E}{v_F} \left( \psi_A^2 - \psi_B^2 \right).
\end{align}
\end{widetext}
Integrating the first equation around $x=0$ and assuming that the spinor only has a finite discontinuity yields
\begin{equation}
    \lim_{\epsilon \rightarrow0} \left. \psi_A \psi_B \right|_{-\epsilon}^{+\epsilon} = 0,
\end{equation}
such that $\psi_A \psi_B$ remains continuous. We can then safely divide the second equation by $\psi_A \psi_B$ and integrate,
\begin{equation}
    \lim_{\epsilon \rightarrow0} \left. \ln \left( \frac{\psi_A}{\psi_B} \right) \right|_{-\epsilon}^{+\epsilon} = 2\alpha.
\end{equation}
Up to a sign we then find
\begin{equation}
    \psi(0^+) = e^{\alpha \sigma_z} \psi(0^-).
\end{equation}

A similar derivation can be performed for an electrostatic delta barrier $v_F \gamma \delta(x) \sigma_0$. In that case, one finds that $\psi_A^2 - \psi_B^2$ remains continuous and
\begin{equation}
    \lim_{\epsilon \rightarrow0} \left. \ln \left( \frac{\psi_A - \psi_B}{\psi_A + \psi_B} \right) \right|_{-\epsilon}^{+\epsilon} = 2 i \tau \gamma,
\end{equation}
yielding
\begin{equation}
    \psi(0^+) = e^{-i\tau \gamma \sigma_x} \psi(0^-).
\end{equation}

\section{Two-terminal transport}

\subsection{Mirror symmetric solution}

In order to make the mirror symmetry ($x \mapsto -x, \sigma_y$) manifest, the solution on the left-hand side
\begin{equation}
    \psi_{s,L}(x) = \frac{a_L}{\sqrt{2}} \begin{pmatrix} 1 \\ \tau e^{is\tau \phi} \end{pmatrix} e^{isqx} + \frac{b_L}{\sqrt{2}} \begin{pmatrix} 1 \\ -\tau e^{-is\tau \phi} \end{pmatrix} e^{-isqx},
\end{equation}
constrains the solution on the right-hand side:
\begin{equation}
    \psi_{s,R}(x) = \frac{a_R}{\sqrt{2}} \begin{pmatrix} i\tau e^{-is\tau \phi} \\ i \end{pmatrix} e^{isqx} + \frac{b_R}{\sqrt{2}} \begin{pmatrix} -i\tau e^{is\tau \phi} \\ i \end{pmatrix} e^{-isqx},
\end{equation}
which now have both the effective time-reversal symmetry and mirror symmetry built in. The scattering matrix, defined by $\left(b_L, \, a_R \right) = S \left( a_L, \, b_R \right)$, is then given by
\begin{equation}
    S = -e^{i s \tau \phi} \sech(\alpha + i s \tau \phi) \begin{pmatrix} \sinh \alpha & i \tau \cos \phi \\ i \tau \cos \phi & \sinh \alpha \end{pmatrix}.
\end{equation}
For the usual Schr\"odinger equation from standard quantum mechanics, the spinor eigenstate is replaced with $1$ and there is no orbital degree of freedom on which a symmetry may act non-trivially.

\subsection{Mixed delta barrier}

Here we consider the more general case of a pseudogauge delta barrier in combination with an electrostatic delta barrier. While the former may be realized experimentally in a nanoslide, the latter can be realized either with a finger gate. However, this may be difficult to align and is incompatible with STM experiments. So instead we suggest to use a global bottom gate [see Fig.\ 1(a) of the main text] as the metal gates will screen the bottom gate except for the suspended region which naturally aligns with the pseudogauge barrier. Moreover, when the Fermi wave length is much larger than the gated region, we can also approximate the electrostatic potential with a delta function. This gives rise to the following potential in the Dirac Hamiltonian:
\begin{equation}
    \Lambda^\mu(x,k_y) = \frac{1}{e} \left( \gamma \delta^\mu_0 + \tau \alpha \delta^\mu_y \right) \delta(x) + f^\mu(k_y).
\end{equation}
The boundary condition becomes
\begin{equation}
    \psi(0^+) = e^{\alpha \sigma_z - i \tau \gamma \sigma_x} \psi(0^-).
\end{equation}
As described in the main text, we use the boundary conditions to obtain the scattering matrix:
\begin{widetext}
\begin{equation}
    \begin{aligned}
    S(k_y,E) & = \frac{1}{q \cosh \sqrt{\alpha^2 - \gamma^2} + i \left( \gamma k_F + \alpha s \tau k_y \right) \sinh \sqrt{\alpha^2 - \gamma^2} / \sqrt{\alpha^2 - \gamma^2}} \\
    & \times \begin{bmatrix} -\left( q + i s \tau k_y \right) \left( \gamma s \tau k_y / k_F + \alpha \right) \tfrac{\sinh \sqrt{\alpha^2 - \gamma^2}}{\sqrt{\alpha^2 - \gamma^2}} & q \\ q & \left( q - i s \tau k_y \right) \left( \gamma s \tau k_y / k_F + \alpha \right) \tfrac{\sinh \sqrt{\alpha^2 - \gamma^2}}{\sqrt{\alpha^2 - \gamma^2}} \end{bmatrix},
    \end{aligned}
\end{equation}
which is independent of energy. The transmission function becomes
\begin{align}
    T(k_y,E) = |t|^2 = |t'|^2 & = \frac{q^2}{q^2 \cosh^2 \sqrt{\alpha^2 - \gamma^2} + \left( \gamma k_F + \alpha s \tau k_y \right)^2 \tfrac{\sinh^2 \sqrt{\alpha^2 - \gamma^2}}{\left( \alpha^2 - \gamma^2 \right)}} \\
    & = \frac{\cos^2 \phi}{\cos^2 \phi \cosh^2 \sqrt{\alpha^2 - \gamma^2} + \left( \gamma + \alpha s \tau \sin \phi \right)^2 \tfrac{\sinh^2 \sqrt{\alpha^2 - \gamma^2}}{\left( \alpha^2 - \gamma^2 \right)}},
\end{align} 
\end{widetext}
where $q = \sqrt{k_F^2 - k_y^2} = k_F \cos \phi$. Note that this expression depends explicitly on the band and valley index through $s\tau \sin \phi$ only when both $\alpha$ and $\gamma$ are nonzero. At normal incidence,
\begin{equation}
    T(0) = \frac{\alpha^2 - \gamma^2}{\alpha^2 \cosh^2 \sqrt{\alpha^2 - \gamma^2} - \gamma^2}.
\end{equation}
Also note that $r = r' = 0$ for $\sin \sqrt{\gamma^2 - \alpha^2} = 0$. This occurs for $\gamma^2 = (n\pi)^2 + \alpha^2 > \alpha^2$ ($n=1,2,\ldots$) and coincides with a resonance due to bound states merging with the continuum: $E_b \rightarrow \pm v_F k_y$ as shown below. We see that $n=0$ does not produce a resonance because $\lim_{z\rightarrow0}\sin(z)/z = 1$.

To find the energy dispersion of the bound states, we consider the poles of the $S$ matrix under analytical continuation $q \rightarrow is\kappa$ with $\kappa = \sqrt{k_y^2 - (E/v_F)^2} > 0$. We find
\begin{equation}
    \sqrt{k_y^2 - \frac{E^2}{v_F^2}} \, \zeta \cosh \zeta + \left( \frac{\gamma E}{v_F} + \tau k_y \alpha \right) \sinh \zeta = 0,
\end{equation}
with $\zeta = \sqrt{\alpha^2 - \gamma^2}$. This is solved by $E = v k_y$ with
\begin{equation}
    \sqrt{1 - \frac{v^2}{v_F^2}} + \sgn(k_y) \left( \tau \alpha + \frac{\gamma v}{v_F} \right) \frac{\tanh \zeta}{\zeta} = 0,
\end{equation}
whose solutions for $v$ are given in the main text.

Finally, we calculate the two-terminal conductance by integrating over all transverse modes. This results in
\begin{widetext}
\begin{align}
    \frac{G}{G_0} & = \int_{-1}^1 dt \, T(k_F t, E) \\
    & = \frac{\alpha^2 - \gamma^2}{\left( \alpha^2 - \gamma^2 \cosh^2 \zeta \right)^2} \left\{ \alpha^2 - \gamma^2 \cosh^2 \zeta +  \alpha \gamma \ln \left| \frac{\alpha + \gamma}{\alpha-\gamma} \right| \sinh^2 \zeta + \ln \left| \tanh \frac{\zeta}{2} \right| \left( \gamma^2 \cosh \zeta \sinh \zeta + \alpha^2 \tanh \zeta \right) \sinh \zeta \right\},
\end{align}
\end{widetext}
with $G_0$ the ballistic conductance, $\zeta = \sqrt{\alpha^2 - \gamma^2}$, and
\begin{align}
    \lim_{\gamma \rightarrow \pm \alpha} \frac{G}{G_0} & = \frac{1 + \alpha^2 \left( 2 \ln |\alpha| - 1 \right)}{\left( 1 - \alpha^2 \right)^2}, \\
    \lim_{\alpha \rightarrow \pm 1} \lim_{\gamma \rightarrow \pm \alpha} \frac{G}{G_0} & = \frac{1}{2}.
\end{align}

\subsection{Transfer-matrix method}

The key idea behind the transfer-matrix method is that any potential profile can be approximated by a series of piecewise constant potentials:
\begin{equation}
    V(x) = \begin{cases}
    V_1 & x < x_1 \\
    V_2 & x_1 < x < x_2 \\
    \vdots & \vdots \\
    V_{N+1} & x > x_N \\
    \end{cases},
\end{equation}
and similar for a gauge potential with $x_1 = -l/2$ and $x_{N+1} = l/2$. Moreover, provided the intervals of length $l/N$ are much smaller than the Fermi wavelength,
\begin{equation}
    \lambda_F = \frac{2\pi}{k_F} = \frac{2\pi}{\sqrt{\pi |n|}} \approx 35 \text{ nm} \sqrt{\frac{10^{12} \text{ cm}^{-2}}{|n|}},
\end{equation}
the charge carriers experience a smooth profile. Hence the number of regions $N$ does not have to be large in order to obtain converged results. For example, for $l = 50$ nm, we can take $N \sim 20$ which is more than sufficient for the density range corresponding to the Dirac regime. In the following, for convenience, we set $E = 0$ as this only amounts to a global energy shift.

In each region, the potentials are constant and the general solution can be written as
\begin{widetext}
\begin{equation} \label{eq:gensol}
    \begin{aligned}
        \psi_n(x) & = a_n \begin{pmatrix} -V_n / v_F \\ \tau k_n + i \pi_n \end{pmatrix} e^{ik_n x} + b_n \begin{pmatrix} -V_n / v_F \\ -\tau k_n + i \pi_n \end{pmatrix} e^{-ik_n x},
    \end{aligned}
\end{equation}
\end{widetext}
with $\tau = \pm1$ the valley index, $n = 1,\ldots,N+1$ and where $a_n$ and $b_n$ are complex coefficients that are determined from boundary conditions. We further defined
\begin{align}
    k_n & = -\sgn(V_n) \sqrt{(V_n / v_F)^2 - \pi_n^2}, \\
    \pi_n & = k_y + e A_{\text{tot},n,y},
\end{align}
where the sign ensures that $k_1$ and $k_{N+1}$ correspond to right-moving modes for both electron ($V_n < 0$) and hole ($V_n > 0$) doping. Continuity of the wave function at $x = x_n$ gives
\begin{equation}
    \begin{pmatrix} a_{n+1} \\ b_{n+1} \end{pmatrix} = T_n \begin{pmatrix} a_n \\ b_n \end{pmatrix},
\end{equation}
with
\begin{widetext}
    \begin{equation}
        T_n = \begin{bmatrix} \frac{V_n \left( k_{n+1} - i \tau \pi_{n+1} \right) + V_{n+1} \left( k_n + i \tau \pi_n \right)}{2k_{n+1} V_{n+1}} \, e^{i(k_n - k_{n+1}) x_n} & \frac{V_n \left( k_{n+1} - i \tau \pi_{n+1} \right) - V_{n+1} \left( k_n - i \tau \pi_n \right)}{2k_{n+1} V_{n+1}} \, e^{-i(k_n + k_{n+1}) x_n} \\
        \frac{V_n \left( k_{n+1} + i \tau \pi_{n+1} \right) - V_{n+1} \left( k_n + i \tau \pi_n \right)}{2k_{n+1} V_{n+1}} \, e^{i(k_n + k_{n+1}) x_n} & \frac{V_n \left( k_{n+1} + i \tau \pi_{n+1} \right) + V_{n+1} \left( k_n - i \tau \pi_n \right)}{2k_{n+1} V_{n+1}} \, e^{-i(k_n - k_{n+1}) x_n}
        \end{bmatrix}.
    \end{equation}
\end{widetext}
We thus obtain
\begin{equation}
    \begin{pmatrix} a_{N+1} \\ b_{N+1} \end{pmatrix} = T \begin{pmatrix} a_1 \\ b_1 \end{pmatrix},
\end{equation}
where
\begin{equation}
    T = T_N \cdots T_2 T_1,
\end{equation}
is the transfer matrix. We now impose the remaining boundary conditions in the leads. For an incident electron wave from the left-hand side, we require
\begin{alignat}{2}
    a_1 & = \frac{N_i}{\sqrt{v_i}}, \qquad && b_1 = r \, \frac{N_r}{\sqrt{-v_r}}, \\
    a_{N+1} & = t \, \frac{N_t}{\sqrt{v_t}}, \qquad && b_{N+1} = 0,
\end{alignat}
where we properly normalized the current. Here the longitudinal velocities $v_i = -v_r = -v_Fk_1/V_1$, $v_t = -v_Fk_{N+1} / V_{N+1}$, and spinor normalization $N_i = N_r = 1 / (\sqrt{2}|V_1 / v_F|)$ and $N_t = 1 / (\sqrt{2}|V_{N+1} / v_F|)$. We then obtain
\begin{equation}
    \begin{pmatrix} tN_t/\sqrt{v_t} \\ 0 \end{pmatrix} =\begin{pmatrix} T_{11} & T_{12} \\ T_{21} & T_{22} \end{pmatrix} \begin{pmatrix} N_i/\sqrt{v_i} \\ rN_r/\sqrt{-v_r} \end{pmatrix},
\end{equation}
which yields
\begin{equation}
    r = -\frac{T_{21}}{T_{22}}, \qquad
    t = \frac{N_i}{N_t} \sqrt{\frac{v_t}{v_i}} \frac{\det(T)}{T_{22}}.
\end{equation}
Similarly, one can consider an incident mode from the right-hand side, which results in
\begin{equation}
    r' = \frac{T_{12}}{T_{22}}, \qquad t' = \frac{N_t}{N_i} \sqrt{\frac{v_i}{v_t}} \frac{1}{T_{22}}.
\end{equation}
Moreover, in the transverse gauge ($A_{\text{tot},x} = 0$), the effective 1D Hamiltonian $\mathcal H(x,k_y)$ has the antiunitary symmetry $\sigma_z \mathcal K$ within a given $k_y$ and valley sector, which is manifest in our general solution of Eq.\ \eqref{eq:gensol}. Thus, this implies that $t = t'$ and $|r|=|r'|$ resulting in the following special relations
\begin{equation}
    \det(T) = \frac{N_t^2}{N_i^2} \frac{v_i}{v_t} = \frac{k_1 V_1}{k_{N+1} V_{N+1}} > 0, \quad |T_{12}|=|T_{21}|.
\end{equation}
We have checked numerically that all these properties are satisfied. The $S$-matrix becomes
\begin{equation}
    S = \frac{1}{T_{22}} \begin{pmatrix} -T_{21} & \sqrt{\det(T)} \, \\ \sqrt{\det(T)} & T_{12} \end{pmatrix},
\end{equation}
and the reflection and transmission functions are
\begin{equation}
    R = \left| \frac{T_{21}}{T_{22}} \right|^2, \qquad
    T = \frac{\det(T)}{|T_{22}|^2}. 
\end{equation}
One can also study bound states by considering imaginary $k_1$ and $k_{N+1}$. In this case, we have
\begin{equation}
    \begin{pmatrix} b \\ 0 \end{pmatrix} =\begin{pmatrix} T_{11} & T_{12} \\ T_{21} & T_{22} \end{pmatrix} \begin{pmatrix} 0 \\ a \end{pmatrix},
\end{equation}
or
\begin{equation}
    b = T_{12} a, \qquad T_{22} = 0,
\end{equation}
where $a$ is fixed by normalization. This is equivalent to $|r|, |t| \rightarrow \infty$ under analytic continuation, i.e., the poles of the $S$-matrix for imaginary $k_1$ and $k_{N+1}$. 

\begin{figure*}
    \centering
    \includegraphics[width=\linewidth]{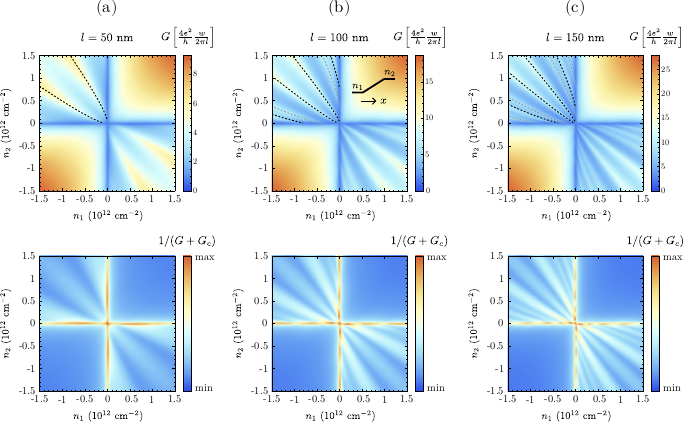}
    \caption{Two-terminal conductance $G(n_1,n_2)$ (top row) and resistance (bottom row) for a graphene nanoslide with $\alpha = 0.8$ as a function of the density in the leads $n_1$ and $n_2$ using a linear density profile. Here the distance between leads $l$ is indicated at the top, and we use the transfer-matrix method with $N = 201$ for the simulation. Dashed lines in the top row correspond to the quantization rule of the main text with $\delta = 0.16$, and where black ($+$) and gray ($-$) correspond to time-reversed phases. For the bottom row, we have included a phenomenological contact resistance $1/G_c$ in parallel with $G_c = 0.1 \times 4e^2 w / (2\pi l)$.}
    \label{fig:G2sm}
\end{figure*}

To model the nanoslide at zero external magnetic field in transverse gauge, we set 
$A_{\text{tot},x} = 0$ and
\begin{equation}
    e l A_{\text{tot},y}(x) = \tau \alpha N \delta_{x,0},
\end{equation}
where $l$ is the distance between the leads and we take $N$ odd. We model the electrostatic potential $V(x) = -v_F \sgn[n(x)] \sqrt{\pi |n(x)|}$ with a linear density profile,
\begin{equation}
    n(x) =
    \begin{cases}
        n_1 & x < -l/2 \\
        \frac{n_1 + n_2}{2} + (n_2 - n_1) \frac{x}{l} & -l/2 < x < l/2 \\
        n_2 & x > l/2
    \end{cases}.
\end{equation}
The two-terminal conductance is then given by
\begin{equation}
    G(n_1,n_2) = \frac{4e^2}{h} \frac{w}{2\pi l} \int_{-k_Fl}^{k_F l} dt \, T( t; n_1, n_2),
\end{equation}
with $t = k_y l$ and $k_F = \sqrt{\pi|n_1|}$. Results for the two-terminal conductance as a function of the density in the leads are shown in Fig.\ \ref{fig:G2sm} using $N = 201$ for $\alpha = 0.8$ and $l = 50, 100$, and $150$~nm. We also show the resistance $1 / (G + G_c)$ with $1 / G_c$ a contact resistance in parallel, where we take $G_c = 0.1 \times 4e^2 w / (2\pi l)$.

\onecolumngrid

\section{Local density of states}

In this section, we show explicitly that the local density of states of the graphene nanoslide in the absence of electrostatic potentials is given by
\begin{equation}
    \frac{\rho(x,E)}{\rho_0} = 1 - |\sinh \alpha| \left[ J_0(2k_Fx) e^{-|\alpha|} + 2 \cosh \alpha \sum_{n=1}^\infty J_{2n}(2k_Fx) e^{-2n|\alpha|} - \cosh \alpha \, e^{-|2k_Fx \sinh \alpha|} \right].
\end{equation}

\subsection{Scattering states}

Here we calculate the contribution from scattering modes to the local density of states. There are two scattering solutions for each spin, valley ($\tau = \pm1$), and band ($s = \pm1$), with energy $E_s = s \sqrt{q^2 - k_y^2}$. These correspond to an incoming wave from the left-hand or right-hand side of the barrier. For $x < 0$ we have
\begin{align}
    |\psi_{s,\tau}^{(L)}|^2 + |\psi_{s,\tau}^{(R)}|^2 & = \frac{1}{2} \left| \begin{pmatrix} 1 \\ \tau e^{is\tau \phi} \end{pmatrix} e^{isqx} + r \begin{pmatrix} 1 \\ -\tau e^{-is\tau \phi} \end{pmatrix} e^{-isqx} \right|^2 + \frac{1}{2}\left| t' \begin{pmatrix} 1 \\ -\tau e^{-is\tau \phi} \end{pmatrix} \right|^2 \\
    & = 2 + \mathrm{Re} \left[ r \left( 1 - e^{-2is\tau\phi} \right) e^{-2isqx} \right] \equiv f_{s,\tau},
\end{align}
where we used $|r|^2 + |t'|^2 = 1$. Taking into account spin, we obtain the following integral
\begin{align}
    & \frac{1}{2\pi^2} \int_0^\infty dq \int_{-\infty}^\infty dk_y \, f_{s,\tau}(q,k_y) \delta \left( E - s v_F \sqrt{q^2 + k_y^2} \, \right) \\
    & = \frac{1}{2\pi^2v_F} \, \Theta(sE) \int_0^\infty dq \int_{-\infty}^\infty dk_y \, f_{s,\tau}(q,k_y) \left| \frac{\sqrt{q^2 + k_y^2}}{q} \right| \delta \left( q - \sqrt{k_F^2 - k_y^2} \, \right) \\
    & = \frac{1}{2\pi^2v_F} \, \Theta(sE) \int_{-k_F}^{k_F} dk_y \, \frac{f_{s,\tau}\left( q = \sqrt{k_F^2 - k_y^2}, k_y \right)}{\sqrt{1 - (k_y/k_F)^2}} \\
    & = \frac{k_F}{2\pi v_F} \, \Theta(sE) \int_{-\pi/2}^{\pi/2} \frac{d\phi}{\pi} \, f_{s,\tau}(\phi),
\end{align}
where we substituted $k_y = k_F \sin \phi$ with $dk_y = d\phi \, k_F \cos\phi$ and used that $\sqrt{1 - (k_y/k_F)^2} = \cos \phi > 0$ in the integration domain. This yields
\begin{align}
    \rho_s(x, E) & = \frac{k_F}{\pi v_F} \sum_{s,\tau=\pm} \Theta(sE) \int_{-\pi/2}^{\pi/2} \frac{d\phi}{\pi} \left\{ 1 + \frac{1}{2} \, \mathrm{Re} \left[ r \left( 1 - e^{-2is\tau\phi} \right) e^{-2isk_F x \cos \phi} \right] \right\} \\
    & = \frac{2k_F}{\pi v_F} \left\{ 1 + \frac{1}{4} \sum_{s,\tau=\pm} \Theta(sE) \int_{-\pi/2}^{\pi/2} \frac{d\phi}{\pi} \, \mathrm{Re} \left[ r \left( 1 - e^{-2is\tau\phi} \right) e^{-2isk_F x \cos \phi} \right] \right\},
\end{align}
with
\begin{equation}
    \frac{\delta \rho_s(x,E)}{\rho_0} = \frac{1}{4} \sum_{s,\tau=\pm} \Theta(sE) \int_{-\pi/2}^{\pi/2} \frac{d\phi}{\pi} \, \mathrm{Re} \left[ r \left( 1 - e^{-2is\tau\phi} \right) e^{-2isk_F x \cos \phi} \right],
\end{equation}
and where $\rho_0 = 2k_F / (\pi v_F)$ is the bulk density of states. In the absence of the electrostatic barrier ($\gamma = 0$), this becomes
\begin{align}
    \frac{\delta \rho_s(x, E)}{\rho_0} & = -\frac{1}{4\pi} \sum_{s,\tau=\pm} \Theta(sE) \int_{-\pi/2}^{\pi/2} d\phi \, \mathrm{Re} \left[ \frac{\sinh \alpha e^{is\tau \phi}}{\cos \phi \cosh \alpha + i s \tau \sin \phi \sinh \alpha} \left( 1 - e^{-2is\tau\phi} \right) e^{-2isk_F x \cos \phi} \right] \\
    & = -\frac{1}{2\pi} \sum_{s,\tau=\pm} \Theta(sE) \int_{-\pi/2}^{\pi/2} d\phi \, \mathrm{Re} \left[ \frac{i s \tau \sinh \alpha \sin \phi}{\cos \phi \cosh \alpha + i s \tau \sin \phi \sinh \alpha} e^{-2isk_F x \cos \phi} \right].
\end{align}
Moreover, since
\begin{equation}
    \frac{1}{\cos \phi \cosh \alpha + i s \tau \sin \phi \sinh \alpha} = \frac{2 \cos \phi \cosh \alpha - 2i s \tau \sin \phi \sinh \alpha}{\cos(2\phi) + \cosh(2\alpha)},
\end{equation}
only the imaginary part contributes because the real part gives an odd integrand. Hence,
\begin{align}
    \frac{\delta \rho_s(x, E)}{\rho_0} & = -\frac{1}{\pi} \sum_{s,\tau=\pm} \Theta(sE) \int_{-\pi/2}^{\pi/2} d\phi \, \frac{\sinh^2 \alpha \sin^2 \phi}{\cos(2\phi) + \cosh(2\alpha)} \cos \left( 2k_F x \cos \phi \right) \\
    & = -\sinh \alpha \int_{-\pi/2}^{\pi/2} \frac{d\phi}{\pi} \, \frac{2\sinh \alpha \sin^2 \phi}{\cos(2\phi) + \cosh(2\alpha)} \cos \left( 2k_F x \cos \phi \right),
\end{align}
as $\Theta(E) + \Theta(-E) = 1$. Now using the Jacobi-Anger identity, specifically
\begin{equation}
    \cos(z \cos \phi) = J_0(z) + 2 \sum_{n=1}^\infty (-1)^n J_{2n}(z) \cos(2n\phi),
\end{equation}
we find
\begin{align}
    \int_{-\pi/2}^{\pi/2} \frac{d\phi}{\pi} \, \frac{2\sinh \alpha \sin^2 \phi}{\cos(2\phi) + \cosh(2\alpha)} & = \sgn(\alpha) e^{-|\alpha|}, \\
    \int_{-\pi/2}^{\pi/2} \frac{d\phi}{\pi} \, \frac{2\sinh \alpha \sin^2 \phi}{\cos(2\phi) + \cosh(2\alpha)} \cos(2n\phi) & = \sgn(\alpha) (-1)^n e^{-2n|\alpha|} \cosh \alpha,
\end{align}
for $n = 1, 2, \ldots$. This result can be obtained using complex analysis:
\begin{equation}
    \int_{-\pi/2}^{\pi/2} \frac{d\phi}{\pi} \, \frac{2 \sin^2 \phi}{\cos(2\phi) + a} e^{i2n\phi} = -\frac{1}{2\pi i} \oint_C dz \, \frac{\left( 1 - z \right)^2}{1 + 2 a z + z^2} z^{n-1}, \label{eq:can1}
\end{equation}
with $a = \cosh(2\alpha) > 1$ and $C$ the complex unit circle. The roots of the denominator are $z_\pm = \pm \sqrt{a^2 - 1} - a$. For $n=0$ there are two zeroth-order poles inside $C$ at $z = 0$ and $z = z_+$ while for $n = 1,2, \ldots$ we only have $z = z_+$. From the residue theorem, we find
\begin{equation}
    -\frac{1}{2\pi i} \oint_C dz \, \frac{\left( 1 - z \right)^2}{1 + 2 a z + z^2} z^{n-1} = -z_+^n \frac{\left( 1 - z_+ \right)^2}{z_+ \left( z_+ - z_- \right)} - \delta_{n,0} = \left( \sqrt{a^2 - 1} - a \right)^n \frac{1 + a}{\sqrt{a^2 - 1}} - \delta_{n,0},
\end{equation}
and plugging in $a = \cosh(2\alpha)$ gives the desired result. We thus obtain
\begin{equation}
    \frac{\delta \rho_s(x, E)}{\rho_0} = - \left| \sinh \alpha \right| \left[ J_0(z) e^{-|\alpha|} + 2 \cosh \alpha \sum_{n=1}^\infty J_{2n}(z) e^{-2n|\alpha|} \right],
\end{equation}
with $z = 2k_F x$.

\subsection{Bound states}

Similarly, we calculate the bound state contribution:
\begin{equation}
    \psi_b^\dag \psi_b = 2a^2 e^{-2\kappa|x|} e^\alpha \cosh \alpha = -e^{2\tau k_y \tanh \alpha \, |x|} \tau k_y \tanh \alpha.
\end{equation}
We find
\begin{align}
    \rho_b(x,E) & = 2 \tanh \alpha \sum_{b,\tau=\pm} \int_{-\infty}^\infty \frac{dk_y}{2\pi} \left(-\tau k_y \right) e^{2\tau k_y \tanh \alpha \, |x|} \delta(E - b v_F k_y \sech \alpha) \Theta(-\tau k_y \alpha) \\
    & = \frac{\sinh \alpha}{\pi v_F} \sum_{b,\tau=\pm} \int_{-\infty}^\infty dk_y \left( -\tau k_y \right)  e^{2\tau k_y \tanh \alpha|x|} \delta\left[ k_y - b E / (v_F \sech \alpha) \right] \Theta(-\tau k_y \alpha),
\end{align}
where we used that $a = \sqrt{\left( 1 - \tanh \alpha \right) \kappa / 2}$ and $\kappa = -\tau k_y \tanh \alpha > 0$. Now substitute $u = -\tau k_y$:
\begin{align}
    \rho_b(x,E) & = \frac{\sinh \alpha}{\pi v_F} \sum_{b,\tau=\pm} \int_{-\infty}^\infty du \, u e^{-2u \tanh \alpha|x|} \delta\left[ u + \tau b E / (v_F \sech \alpha) \right] \Theta(u\alpha) \\
    & = \frac{|\sinh \alpha|}{\pi v_F} \sum_{b,\tau=\pm} \int_0^\infty du \, u e^{-2u |x \tanh \alpha|} \delta\left[ u + \tau b E / (v_F \sech \alpha) \right],
\end{align}
since for $\alpha < 0$ we substitute $u \rightarrow -u$. This finally yields
\begin{align}
    \rho_b(x,E) & = \frac{|\sinh \alpha|}{\pi v_F^2} \cosh \alpha \sum_{b,\tau=\pm} \left( - \tau b E \right) e^{-2\tau b E |x \sinh \alpha|/v_F} \Theta(-\tau bE) \\
    & = \frac{|E \sinh \alpha|}{\pi v_F^2} \cosh \alpha \, e^{-2|E x \sinh \alpha|/v_F} \sum_{b,\tau=\pm} \Theta(-\tau bE) \\
    & = \frac{2k_F}{\pi v_F} |\sinh \alpha| \cosh \alpha \, e^{-|2k_Fx \sinh \alpha|},
\end{align}
since only two terms of the sum contribute regardless of the sign of the energy. This tedious calculation can be prevented by noting that the $x$ dependence is fixed and the density of states has to equal that of a linear chiral dispersion ($E = vk_y$). Taking into account spin and valley degeneracy:
\begin{equation}
    \rho_b(x,E) = \frac{2\kappa}{\pi |v|} \, e^{-2\kappa|x|},
\end{equation}
with $\kappa = k_F \sqrt{(v_F/v)^2 - 1} = k_F |\sinh \alpha|$ and $v = v_F \sech \alpha$.

\subsection{Sublattice polarization}

Here we compute the sublattice-resolved local density of states. Similar as before, we first consider scattering states with $x < 0$. We have
\begin{equation}
    \left[ \psi_{s,\tau}^{(L)} \right]^\dag \sigma_z \psi_{s,\tau}^{(L)} + \left[ \psi_{s,\tau}^{(R)} \right]^\dag \sigma_z \psi_{s,\tau}^{(R)}  = \mathrm{Re} \left[ r \left( 1 + e^{-2is\tau\phi} \right) e^{-2isqx} \right],
\end{equation}
so the only difference is the absence of the background and the sign in the term between parentheses. We thus find
\begin{align}
    \frac{\sigma_s(x,E)}{\rho_0} & = \frac{1}{4} \sum_{s,\tau=\pm} \Theta(sE) \int_{-\pi/2}^{\pi/2} \frac{d\phi}{\pi} \, \mathrm{Re} \left[ r \left( 1 + e^{-2is\tau\phi} \right) e^{-2isqx} \right] \\
    & = -\sinh(\alpha) \int_{-\pi/2}^{\pi/2} \frac{d\phi}{\pi} \, \frac{2 \cosh \alpha \cos^2 \phi}{\cos(2\phi) + \cosh(2\alpha)} \cos \left( 2k_F x \cos \phi \right).
\end{align}
We find
\begin{align}
    \int_{-\pi/2}^{\pi/2} \frac{d\phi}{\pi} \, \frac{2\cosh \alpha \cos^2 \phi}{\cos(2\phi) + \cosh(2\alpha)} & = e^{-|\alpha|}, \\
    \int_{-\pi/2}^{\pi/2} \frac{d\phi}{\pi} \, \frac{2\cosh \alpha \cos^2 \phi}{\cos(2\phi) + \cosh(2\alpha)} \cos(2n\phi) & = -(-1)^n e^{-2n|\alpha|} |\sinh \alpha|,
\end{align}
for $n = 1, 2, \ldots$. Similar as before, this result can be obtained using complex analysis:
\begin{equation}
    \int_{-\pi/2}^{\pi/2} \frac{d\phi}{\pi} \, \frac{2 \cos 2 \phi}{\cos(2\phi) + a} e^{i2n\phi} = \frac{1}{2\pi i} \oint_C dz \, \frac{\left( 1 + z \right)^2}{1 + 2 a z + z^2} z^{n-1}
\end{equation}
with $a = \cosh(2\alpha) > 1$ and $C$ the complex unit circle. This has the same pole structure as Eq.\ \eqref{eq:can1}. From the residue theorem, we find
\begin{equation}
    \frac{1}{2\pi i} \oint_C dz \, \frac{\left( 1 + z \right)^2}{1 + 2 a z + z^2} z^{n-1} = \delta_{n,0} + z_+^n \frac{\left( 1 + z_+ \right)^2}{z_+ \left( z_+ - z_- \right)} = \delta_{n,0} + \left( \sqrt{a^2 - 1} - a \right)^n \frac{1 - a}{\sqrt{a^2 - 1}},
\end{equation}
and plugging in $a = \cosh(2\alpha)$ gives the desired result. This calculation was for $x < 0$. Using the mirror symmetry, we finally obtain for all $x$:
\begin{equation}
    \frac{\sigma_s(x, E)}{\rho_0} = \sgn(x) \sinh \alpha \left[ J_0(z) e^{-|\alpha|} - 2 |\sinh \alpha| \sum_{n=1}^\infty J_{2n}(z) e^{-2n|\alpha|} \right],
\end{equation}
with $z = 2k_F x$ and which is also odd in $\alpha$ as expected. For the bound states, we have
\begin{equation}
    \psi_b^\dag \sigma_z \psi_b = \sgn(x) 2a^2 e^{-2\kappa|x|} e^\alpha \sinh \alpha = -\sgn(x) e^{2\tau k_y \tanh \alpha \, |x|} \tau k_y \tanh^2 \alpha.
\end{equation}
Similar as before, we obtain
\begin{equation}
    \sigma_b(x,E) = \sgn(x) \, \frac{2k_F}{\pi v_F} |\sinh \alpha | \sinh \alpha \, e^{-|2k_F x \sinh \alpha|}.
\end{equation}

\subsection{Numerical calculation}

\subsubsection{Scattering states}

In general, in the presence of both pseudogauge ($\alpha$) and electrostatic ($\gamma$) barriers, using the expression for the reflection coefficient $r$, we have for $x<0$:
\begin{equation}
    \frac{\delta \rho_{s,\pm}(x,E)}{\rho_0} = \int_{-\pi/2}^{\pi/2} \frac{d\phi}{\pi} \, \mathrm{Re}\left[ f_\pm(\phi) e^{-i\sgn(E)2k_Fx\cos(\phi)} \right],
\end{equation}
and thus from the mirror symmetry ($x \mapsto -x, \sigma_y$), which we also verified explicitly using $r'$, we find
\begin{align}
    \frac{\delta \rho_{s,+}(x,E)}{\rho_0} & = \int_{-\pi/2}^{\pi/2} \frac{d\phi}{\pi} \, \mathrm{Re}\left[ f_+(\phi) e^{i\sgn(E)|2k_Fx|\cos(\phi)} \right], \\
    \frac{\delta \rho_{s,-}(x,E)}{\rho_0} & = \sgn(x) \int_{-\pi/2}^{\pi/2} \frac{d\phi}{\pi} \, \mathrm{Re}\left[ f_-(\phi) e^{i\sgn(E)|2k_Fx|\cos(\phi)} \right],
\end{align}
with
\begin{align}
    f_+(\phi) & = -\frac{\zeta \sin^2(\phi) \sinh \zeta \left( \zeta \sinh \zeta + i \gamma \cos \phi \cosh \alpha \right)}{\zeta^2 \cos^2(\phi) \cosh^2(\alpha) + 2i\gamma \zeta \cos \phi \cosh \alpha \sinh \zeta + \left[ \alpha^2 \sin^2(\phi) - \gamma^2 \right] \sinh^2(\zeta)}, \\
    f_-(\phi) & = -\frac{\alpha \cos^2(\phi) \sinh \zeta \left( \zeta \cosh \alpha + i \gamma \cos \phi \sinh \zeta \right)}{\zeta^2 \cos^2(\phi) \cosh^2(\alpha) + 2i\gamma \zeta \cos \phi \cosh \alpha \sinh \zeta + \left[ \alpha^2 \sin^2(\phi) - \gamma^2 \right] \sinh^2(\zeta)}.
\end{align}
Here, the imaginary parts are proportional to $\gamma$ and break electron-hole symmetry.

\subsubsection{Bound states}

The bound-state contribution for the case $\alpha^2 > \gamma^2$ can be written as
\begin{equation}
    \rho_{b,+}(x,E) = \sum_{b=\pm} \Theta(-\alpha E / v_b) \frac{2\kappa_b}{\pi|v_b|} \, e^{-2\kappa_b|x|},
\end{equation}
where we already accounted for the valleys with
\begin{equation}
    \kappa_\pm = k_F \sqrt{(v_F / v_\pm)^2 - 1}, \qquad \frac{v_\pm}{v_F} = \frac{\pm \alpha \sech \zeta - \gamma}{\alpha \mp \gamma \sech \zeta},
\end{equation}
and we used that $E_b = v_b k_y$ for bound states. This follows from the fact that each semi-infinite branch has to contribute a density of states given by $2 / (\pi|v_b|)$ when summing over valleys. On the other hand, for $\alpha^2 < \gamma^2$ we replace
\begin{equation}
    \Theta(-\alpha E / v_b) \rightarrow \Theta \left( \left[ \alpha \cot(-i\zeta) + b \gamma \csc(i\zeta) \right] E / v_b \right).
\end{equation}

\subsection{Electrostatic delta barrier}

Here, we consider the case $\alpha = 0$. We find that
\begin{align}
    f_+(\phi) & = \frac{\sin \gamma \sin^2 \phi}{i \cos \phi - \sin \gamma}, \\
    f_-(\phi) & = 0.
\end{align}
Hence, we have to solve the following integral (for $x< 0$),
\begin{equation}
     \mathrm{Re} \int_{-\pi/2}^{\pi/2} \frac{d\phi}{\pi} \, \frac{\sin \gamma \sin^2 \phi}{i \cos \phi - \sin \gamma} \, e^{iu\cos \phi} = \sum_{n=-\infty}^\infty J_n(u) \, \mathrm{Re} \int_{-\pi/2}^{\pi/2} \frac{d\phi}{\pi} \left( -\frac{a^2 \sin^2}{a^2 + \cos^2 \phi} - i \frac{a \cos \phi \sin^2}{a^2 + \cos^2 \phi} \right) e^{in\left(\phi + \pi/2 \right)},
\end{equation}
with $u = -\sgn(E) 2k_F x$, $-1 \leq a = \sin \gamma \leq 1$ and where we used the Jacobi-Anger identity. Note that the first term in parentheses only has contributions from even $n=2k$ while the second term only has contributions from odd $n=2k+1$. We obtain the following two integrals:
\begin{equation}
    I_{1,k} = -a^2 (-1)^k \int_{-\pi/2}^{\pi/2} \frac{d\phi}{\pi} \frac{\sin^2 \phi}{a^2 + \cos^2 \phi} \, e^{i2k\phi}, \qquad I_{2,k} = a (-1)^k \int_{-\pi/2}^{\pi/2} \frac{d\phi}{\pi} \frac{\cos \phi \sin^2 \phi \, e^{i\phi}}{a^2 + \cos^2 \phi} \, e^{i2k\phi}.
\end{equation}
Substituting $z = e^{i2\phi}$ gives
\begin{equation}
    I_{1,k} = \frac{a^2(-1)^k }{2\pi i} \oint_C dz \, \frac{\left( 1 - z \right)^2}{z^2 + 2 \left( 2a^2 + 1 \right) z + 1} z^{n-1} =
    \begin{cases}
        -\frac{1}{2} \left( 1 + z_+ \right), & \qquad k = 0; \\
        -|a| \sqrt{1 + a^2} \left( -z_+ \right)^{|k|}, & \qquad k = \pm1, \pm2, \ldots;
    \end{cases}
\end{equation}
with $z_+ = 2|a| \sqrt{1 + a^2} - 1 - 2a^2$. Similarly, for the other integral, we find
\begin{equation}
    I_{2,k} = -\frac{a(-1)^k }{4\pi i} \oint_C dz \, \frac{\left( 1 - z \right)^2 \left( 1 + z \right)}{z^2 + 2 \left( 2a^2 + 1 \right) z + 1} z^{n-1} = (-1)^k
    \begin{cases}
        -\frac{a z_+}{2}, & \qquad k = 0,-1; \\
        \frac{a}{2} \left( 1 - z_+ \right) z_+^{|k| - \Theta(-k)}, & \qquad k = 1, \pm2, \pm3, \ldots;
    \end{cases}
\end{equation}
Using that $J_{-n}(z) = (-1)^n J_n(z)$ we find
\begin{align}
    \frac{\delta \rho_s(x,E)}{\rho_0} & = I_{1,0} J_0(z) + 2 \sum_{k=1}^\infty I_{1,k} J_{2k}(z) + 2 \sgn(E) \left[ I_{2,0} J_1(z) + \sum_{k=1}^\infty I_{2,k} J_{2k+1}(z) \right], \\
    & = -\frac{1 + z_+}{2} J_0(z) - 2 |a| \sqrt{1 + a^2} \sum_{k=1}^\infty \left( -z_+ \right)^k J_{2k}(z) \\
    & + a \sgn(E) \left[ -z_+ J_1(z) + (1 - z_+) \sum_{k=1}^\infty \left( -z_+ \right)^k J_{2k+1}(z) \right],
\end{align}
with $z = |2k_F x|$ and $a = \sin \gamma$. These series converge because $|z_+| < 1$. We have numerically verified that our result is correct. Note that, contrary to Ref.\ \cite{yokoyama_gate-controlled_2010}, we find that the scattering-state contribution also contains a piece that is odd in the energy, which is allowed by symmetry. However, the odd piece vanishes at the origin, such that $\rho(x=0,E) - \rho(x=0,-E) = \rho_b(x=0,E) - \rho_b(x=0,-E)$ for $\alpha = 0$. The bound-state contribution becomes
\begin{equation}
    \rho_b(x,E) = \Theta(-E \tan \gamma) \frac{4\kappa}{\pi |v|} \, e^{-2\kappa|x|},
\end{equation}
with $\kappa = k_F \sqrt{(v_F/v)^2 - 1}$ and $v = v_F \cos \gamma$. We thus find
\begin{equation}
    \frac{\delta \rho(x=0,E)}{\rho_0} = I_{1,0} + 2 \Theta(-E \tan \gamma) \frac{2\kappa}{k_F |v/v_F|} = \sin^2 \gamma - |\sin \gamma| \sqrt{1 + \sin^2 \gamma} + 2 \Theta(-E \tan \gamma) | \tan \gamma \sec \gamma |,
\end{equation}
where in general, we have
\begin{equation}
    I_{1,0} = \int_{-\pi/2}^{\pi/2} \frac{d\phi}{\pi} \, \mathrm{Re} \, f_+(\phi).
\end{equation}

\section{Tight-binding simulations}




Figures 2 and 4(c)--4(e) of the main text are based on standard tight-binding models describing electrons hopping between nearest-neighbor carbon $p_z$ orbitals:
\begin{equation} \label{eq:Htb}
    H = \sum\limits_{\left\langle {m,n} \right\rangle } t_{mn} c_m^\dagger  c_n + \sum\limits_n U_n c_n^\dagger c_n\ ,
\end{equation}
where $c_n^\dag$ ($c_n$) creates (destroys) an electron on site $n$ at position $\mathbf{r}_n$. The first sum in Eq.\ \eqref{eq:Htb} runs over pairs of carbon atoms that are nearest to each other, such that $|\mathbf{r}_m-\mathbf{r}_n|=a_0\approx 0.142\unit{nm}$ in the absence of strain, and the second sum describes the onsite energy $U_n=U(\mathbf{r}_n)$. To account for the effect of strain, we adopt the central-force approximation using a phenomenological hopping function given by \cite{pereira_tight-binding_2009}
\begin{equation}
    t_{mn} = -t_0 \exp\left[ -\beta \left( \frac{|\mathbf r_m - \mathbf r_n|}{a_0} - 1 \right) \right],
\label{eq tij}
\end{equation}
where $t_0 = 3\unit{eV}$ is the nearest-neighbor hopping parameter, and we take $\beta = 3.37$.

Furthermore, we assume the translational invariance along the nanoslide ($y$-axis of Fig.\ 1 of the main text) and adopt the periodic boundary hopping (see Ref.\ \cite{chakraborti_electron_2024}) which carries the Bloch momentum $k_y$ via the Bloch phase $e^{ik_y w}$ with the periodicity $w=\sqrt{3}a_0$ ($w=3 a_0$) for the case of armchair (zigzag) along the $x$-axis. The Hamiltonian \eqref{eq:Htb} is then modified to $H\longrightarrow H(k_y)$, and all quantities based on the Green's function,
\begin{equation}
G^r(E,k_y) = \frac{1}{E-H(k_y)+\Sigma_1(E,k_y)+\Sigma_2(E,k_y)}
\label{eq Gr(E,ky)}
\end{equation}
become $k_y$-dependent, such as the transmission function
\begin{equation}
T(k_y) = \mathrm{Tr}\left[\Gamma_1(k_y) G^r(k_y)\Gamma_2(k_y) G^a(k_y)\right]
\label{eq T(ky)}
\end{equation}
and the local density of states (at site $n$)
\begin{equation}
\rho(x_n,k_y) = -\frac{1}{\pi}\func{Im}G^r_{n,n}(k_y)\ .
\label{eq rho_n(ky)}
\end{equation}
Apart from a normalization factor, the conductance shown in Figure 2 of the main text is $G\propto\int_{-\pi}^\pi T(k_y) d k_y w$, and the local densities of states shown in Figure 4 of the main text are based on $\rho_A(x)\propto\int_{-\pi}^\pi \rho(x_n,k_y)dk_y w$, where $w=\sqrt{3}a_0$ is the graphene lattice periodicity along the zigzag direction parallel to the $y$-axis.

\end{document}